\documentclass[prb,twocolumn,nopacs,nofootinbib]{revtex4-2}

\usepackage{amsmath}
\usepackage[dvipsnames]{xcolor}
\usepackage{nicefrac}
\usepackage{graphicx}
\usepackage{microtype}
\usepackage{physics}
\usepackage{mathrsfs}
\usepackage{amssymb}
\usepackage{bbold}
\usepackage{ dsfont }
\usepackage{pifont}
\usepackage{multirow}
\usepackage[bookmarks=true,colorlinks,linkcolor=blue,urlcolor=blue,citecolor=blue]{hyperref}

\usepackage[english]{babel}
\usepackage[applemac]{inputenc}
\usepackage[T1]{fontenc}

\newcommand{\beq}{\begin{equation}}
\newcommand{\eeq}{\end{equation}}
\newcommand{\beqa}{\begin{eqnarray}}
\newcommand{\eeqa}{\end{eqnarray}}

\newcommand{\cmark}{\ding{51}}%
\newcommand{\xmark}{\ding{55}}

\usepackage{soul}
\usepackage[normalem]{ulem}

\graphicspath{{../}}

\begin{document}

\title{Robust quantum boomerang effect in non-Hermitian systems}
\author{Flavio Noronha$^{1,2}$, Jos\'e A. S. Louren\c{c}o$^{1}$ and Tommaso Macr\`{i}$^{3,1}$}
\affiliation{
$^{1}$Departamento de F\'{i}sica Te\'{o}rica e Experimental, Universidade Federal do Rio Grande do Norte, Campus Universit\'{a}rio, Lagoa Nova, Natal-RN 59078-970, Brazil\\
$^{2}$Institute for Theoretical Physics, Utrecht University, Princetonplein 5, 3584 CC Utrecht, The Netherlands\\
$^{3}$ITAMP, Harvard-Smithsonian Center for Astrophysics, Cambridge, Massachusetts 02138, USA
}

\date{\today}

\begin{abstract}
Anderson localization is a general phenomenon that applies to a variety of disordered physical systems.
Recently, a novel manifestation of Anderson localization for wave packets launched with a finite average velocity was proposed, the Quantum boomerang effect (QBE).
This phenomenon predicts that the disorder-averaged center of mass of a particle initially moves ballistically, then makes a U-turn, and finally slowly returns to its initial position. The QBE has been predicted to take place in several Hermitian models with Anderson localization and has been experimentally observed in the paradigmatic quantum kicked rotor model. 
In this work, we investigate the emergence of the QBE in non-Hermitian systems and clarify the importance of symmetries of the Hamiltonian and the initial state. 
We generalize the analytical arguments available in the literature and 
show that even in the case of complex spectrum a boomerang-like behavior can appear in a non-Hermitian system.
We confirm our analytical results through a careful numerical investigation of the dynamics for several non-Hermitian models.
We find that non-Hermiticity leads to the breakdown of the dynamical relation, though the QBE is preserved.
This work opens up new avenues for future investigations in Anderson localized systems. 
The models studied here may be implemented using cold atoms in optical lattices.
\end{abstract}

\maketitle

\section{Introduction}
\label{sec:intro}
It is well known that a propagating wave in a disordered or random environment is subject to multiple scatterings and that may lead to destructive self-interference. This interference is the cause of several phenomena in nature, including the Anderson localization (AL) of quantum particles, i.e.~the absence of wave diffusion in disordered potentials~\cite{Anderson1958Absence}. AL has been experimentally observed in many platforms, including light~\cite{Chabanov2000,Schwartz2007}, ultrasound waves~\cite{Hu2008} and atomic matter~\cite{Julien2008,Manai2015,Billy2008,Jendrzejewski2012,Semeghini2015}. AL is not restricted to Hermitian models. There are several non-Hermitian models in which the disorder leads to this localization, including, e.g.~models with asymmetric hopping~\cite{HatanoNelson1996Localization,HatanoNelson1997Vortex,HatanoNelson1998Non-Hermitian,Efetov1997Directed,Feinberg1997Non-Hermitian,Feinberg99Non-Hermitian,Brouwer97Theory,Goldsheid98Distribution,Nelson98Non-Hermitian,HatanoNelson2016Non-Hermitian,Mudry98Random,Fukui98Breakdown,Yurkevich1999Delocalization,Longhi2015Robust,Zeng2017Anderson,McDonald2018Phase-Dependent,Gong2018,Hamazaki2019Non-Hermitian,Jiang2019Interplay,Zeng2020Topological,Kawabata2021Nounitary,Scriffer2021Anderson} and models with gain and loss parameters, i.e.~complex onsite potentials~\cite{Freilikher1994Effect,Beenakker1996Probability,Paasschens1996localization,Bruce1996Multiple,Longhi2019Topological,Tzortzakakis2020non-Hermitian,Huang2020Anderson}, that can be experimentally implemented with several platforms~\cite{Longhi2015Robust,Rivet2018Constant,Weidemann2022Topological}.

Interesting transitions can appear in non-Hermitian systems as one changes the parameters of the model. These include not only the Anderson transition, but also topological transitions~\cite{Scriffer2021Anderson,Weidemann2022Topological,Longhi2019} and transitions from a phase with complex eigenenergies to a phase where all eigenenergies are real, called real-complex transitions. 
The real-complex transitions usually occur, but are not restricted to, in models with parity-time (\textit{PT}) symmetry~\cite{Bender1998,Bender_2015,Benderbook2019,Lourenco2018,Lourenco2021,Ygor2021,Diego2021,Longhi2019,Zhang2022,Yang2022arxiv} and models with time-reversal (\textit{T}) symmetry~\cite{Gong2018,Longhi2021}. 
%

The non-Hermitian Hatano-Nelson (HN) and the Hermitian Aubry-Andr\'e (AA) models are paradigmatic examples showing AL transitions.
The HN model describes a disordered system with asymmetric hoppings~\cite{HatanoNelson1996Localization}. 
Interestingly, the three transitions (the Anderson, the topological and the real-complex transitions) take place simultaneously in the HN model with periodic boundary conditions. 
As one increases the non-Hermiticity in that model, the system moves from a phase with all states localized, all eigenenergies real and a trivial topological number to a phase with some extended states, complex eigenenergies and a non-trivial topological number~\cite{Hamazaki2019Non-Hermitian,Gong2018}. The AA model is a well-known Hermitian model with pseudo-random potential which possesses a mobility edge~\cite{Aubry1980,Sinai1987}. 
There are several non-Hermitian generalizations of the AA model which also display localization transitions~\cite{Zeng2017Anderson,Zeng2020Topological,Liu2020Generalized,Longhi2019,Longhi2021}. 
In Ref.~\cite{Scriffer2021Anderson} the authors investigate a \textit{PT}-symmetric non-Hermitian AA model and find three distinct phases. In the \textit{PT}-broken phase the system presents complex eigenenergies and extended states. There are two \textit{PT}-unbroken phases in the model. In one of them there are extended states while in the other one all states are localized. 

AL has direct implications for the transport properties of disordered materials. A genuine dynamical phenomenon due to AL, called quantum boomerang effect (QBE) has been recently introduced in Ref.~\cite{Prat2019Quantum}. 
In order to explain the QBE it is pertinent to recall the behavior of a traditional boomerang. 
Releasing a boomerang with an initial velocity, the boomerang will initially move away from its origin and, after some time, return to its initial position. 
Analogously, in the quantum realm, releasing a particle launched with momentum $k_0$ in a \textit{T}-symmetric disordered potential leads to the QBE: The disorder-averaged center of mass (DACM) of the particle $\overline{\langle x(t)\rangle}$ initially moves balistically departing from the origin [i.e.~$\overline{\langle x(t)\rangle}\approx v_0t$ for $t\ll \tau$, where $\tau$ is the mean free time], then makes a U-turn, and slowly returns to its initial position, $\overline{\langle x(+\infty)\rangle}=0$~\cite{Prat2019Quantum}. Here $\overline{(\cdots)}$ denotes disorder average, $\langle \cdots \rangle$ is the expectation value of an observable and $v_0$ is the initial velocity of the wave packet. In practice, when the QBE is present one can observe $\overline{\langle x(t)\rangle}\approx 0$ at finite times (if $t\gg \tau$) even in the limit of large system sizes.
This intriguing quantum phenomenon is different from the behavior expected for the classical disordered counterpart, where the center of mass would initially move away from the origin and saturate at an approximate distance $\ell$ from the starting point, where $\ell$ is the mean free path~\cite{Prat2019Quantum}.
The QBE was predicted to occur in several models with AL~\cite{Tessieri2021Quantum}. 
Most of the models satisfy the dynamical relation~\cite{Prat2019Quantum,Tessieri2021Quantum}
\begin{equation}
    \frac{d}{dt}\overline{\langle x^2(t)\rangle}=2v_0 \overline{\langle x(t)\rangle},\label{dynamical}
\end{equation}
which shows that the DACM is connected to the spreading of the wave packet. According to it, $\overline{\langle x(t)\rangle}$ can only come back to the origin when the wave function stops spreading.
Recently the QBE was verified in an experimental realization of the quantum kicked rotor~\cite{Sajjad2021Observation}.


All the previous results on the QBE were found in Hermitian \textit{T}-symmetric systems and several questions arose concerning which are the most general initial conditions and properties of the model required for the QBE to exist. 
Very recently two independent works established that \textit{T} symmetry of the Hamiltonian is not fundamental to observe the QBE and showed sufficient conditions to observe the effect~\cite{Janarek2022,Noronha2022}. 
Though~\cite{Noronha2022} focused in Hermitian models, the presented analytical arguments predict the QBE in both Hermitian and non-Hermitian systems with real spectrum.

In the present manuscript we numerically confirm the presence of the QBE in several non-Hermitian models, establishing that Hermiticity is not a necessary condition for the QBE to exist. Here we also investigate the main characteristics of this effect in non-Hermitian systems and show models where the QBE is not observed.
We focus on general one-dimensional disordered lattice models which display AL. We study their dynamics employing extensive numerical simulations. 

We briefly summarize the main findings and the structure of the paper.
In Sec.~\ref{analytical} we review and generalize the analytical arguments that lead us to expect a boomerang behavior in both Hermitian and non-Hermitian systems.
In Sec.~\ref{model} we discuss numerical results showing the QBE in the HN model and in a random-hopping model, which are non-Hermitian models with \textit{T} symmetry. 
If the model is \textit{T} symmetric a boomerang-like behavior can be present even in the case of complex eigenenergies.
We also show the presence of the QBE in the Hermitian AA model and in its non-Hermitian \textit{PT}-symmetric generalization. The QBE is also shown in a non-Hermitian random hopping model with disorder in the complex phase of the hopping. This model is more general than all the other models considered in the literature about QBE in the sense that it simultaneously breaks Hermiticity, \textit{T}, \textit{P} and \textit{PT} symmetry.
Investigating other models without \textit{T} or \textit{PT} symmetry in phases with AL and complex spectrum we find that the DACM, instead of reaching the origin, typically presents a local minimum after the U-turn, and then increases again. This is the case of e.g.~the HN model with complex potential, the Anderson model with alternating gain and loss with constant magnitude, and a random gain and loss model. 
In Section~\ref{conclusion} we summarize our findings, mention possible experimental realizations of some non-Hermitian models and provide interesting scenarios to generalize this work. 
Finally, in the Appendix we discuss some details regarding the saturation of the variance when the QBE is present and the breakdown of the dynamical relation in the non-Hermitian models considered here.

\section{Derivation of the QBE}\label{analytical}
In this Section we provide a review of the analytical arguments with sufficient conditions for the existence of the QBE in non-Hermitian models, as discussed in Ref.~\cite{Noronha2022}, and extend the discussion for the boomerang-like behavior in the case of complex eigenenergies. We focus on the symmetries of the initial state and the disordered non-Hermitian Hamiltonian.
For compactness of notation, we discuss one-dimensional models. However, all the considerations below can be immediately generalized to an arbitrary number of spatial dimensions. 
The boomerang effect is expected to appear if {\it (a)} the Hamiltonian presents AL, {\it (b)} 
the spectrum is real or, in the case of complex spectrum, the Hamiltonian $H$ is \textit{T} symmetric,
{\it (c)}~the ensemble $\{H\}$ of all disorder realizations of the model is \textit{PT} invariant, $\mathcal{PT}\{H\}(\mathcal{PT})^{-1}= \{H\}$, i.e.~for each disorder realization $H$, its parity-time counterpart $\tilde{H}={\mathcal{PT}} H({\mathcal{PT}})^{-1}$ is also a disorder realization of the same model,
and {\it (d)} the initial state is an eigenstate of $\mathcal{PT}$, $\mathcal{PT} |\psi_0 \rangle=\pm |\psi_0 \rangle$~\cite{Noronha2022}. Here $\mathcal{T}$ is the time-reversal operator and $\mathcal{P}$ is the spatial inversion operator.


In order to demonstrate the presence of the boomerang effect with the previous conditions, in the following we write expressions for the center of mass $\langle x(t)\rangle$. 
We divide the demonstration in the two different cases of condition \textit{(b)}. In the case where all eigenvalues are real, expanding $\ket{\psi_0}=\sum_n c_n \ket{\phi_n}$ in terms of the eigenvectors of the Hamiltonian, $H\ket{\phi_n}=\epsilon_n \ket{\phi_n}$, using condition {\it (a)}, taking the average over disorder realizations and taking the limit $\overline{\langle x(+\infty)\rangle}:=\lim_{t\to+\infty}\overline{\langle x(t)\rangle}$ one finds the diagonal ensemble~\cite{Prat2019Quantum,Sajjad2021Observation,Janarek2022,Noronha2022,footeq2}
\begin{eqnarray}
\overline{\langle x(+\infty)\rangle}
&=&\overline{  
\sum_{n}
\left|\braket{\psi_0}{\phi_{n}}\right|^2 \bra{\phi_n}X \ket{\phi_n}
},\label{temp1}
\end{eqnarray}
where $X$ is the position operator. 
This diagonal ensemble is a good approximation at finite times (if $t\gg \tau$) even in the limit of large system sizes.
Following similar steps, we find the same expression for $\overline{\langle x(-\infty)\rangle}$ and then
%
%
\begin{eqnarray}
\overline{\langle x (+\infty)\rangle} &=& \overline{\langle x (-\infty)\rangle}.\label{eqdiag2}
\end{eqnarray}

Equations~(\ref{temp1}) and (\ref{eqdiag2}) are valid for real spectrum. 
Now we consider the case of complex eigenenergies. Let $|\phi_l\rangle$ be the eigenstate with finite overlap with $|\psi_0\rangle$ ($\bra{\phi_{l}}\ket{\psi_0}\neq0$) which has the largest value for the imaginary part of its eigenvalue, i.e.~$\varepsilon_n''< \varepsilon_{l}'', \, \forall \,n$.
Similarly, let $|\phi_m\rangle$ be the eigenstate which has the smallest value for the imaginary part of its eigenvalue, i.e.~$\varepsilon_{m}''< \varepsilon_n'', \, \forall \,n$. For simplicity we assume that $\epsilon_l''$ and $\epsilon_m''$ are non-degenerate.
Once there are complex eigenenergies the time-evolution of the wave function is not norm preserving. 
The center of mass is given by $\langle x(t)\rangle=\langle \psi(t)|\, X\, |\psi(t)\rangle/\langle \psi(t) | \psi(t)\rangle$, where $| \psi(t)\rangle=\textrm{exp}(-i H t) |\psi_0\rangle$ and $\langle \psi(t)|=\langle \psi_0| \textrm{exp}(+i H^\dagger t) $. 
%
Inserting the identity $I=\sum \ket{\phi_n}\bra{\phi_n}$ in the expression for $\langle x(t)\rangle$ and using the previous relations we find, for $t\to+\infty$ and $t\to-\infty$, 
\begin{eqnarray}
\langle x (+\infty)\rangle&=&\langle \phi_{l}|\,X\,|\phi_{l}\rangle ,\label{eq.t+}\\
\langle x (-\infty)\rangle&=&\langle \phi_{m}|\,X\,|\phi_{m}\rangle.\label{eq.t-}
\end{eqnarray}

We recall that if the Hamiltonian is \textit{T} (or \textit{PT}) symmetric and has a complex eigenvalue $\varepsilon$, then $\varepsilon^*$ is also an eigenvalue. Indeed, if $K=\mathcal{T}$ (or $K=\mathcal{PT}$) and $\ket{\phi}$ is an eigenvector of $H$, then
\begin{eqnarray}
 H \ket{\phi}&=&\varepsilon\ket{\phi},\nonumber\\
 KH \ket{\phi}&=&H(K \ket{\phi})=K \varepsilon\ket{\phi}=\varepsilon^*(K\ket{\phi}),
\end{eqnarray}
and $K\ket{\phi}$ is the associated eigenvector.
%
For a \textit{T}-symmetric $H$ we use the property above to find $\varepsilon_{m}=\varepsilon_{l}^*$, $|\phi_{m}\rangle =\mathcal{T}|\phi_{l}\rangle$ and  $\phi_{m}(x)=\phi_{l}^*(x)$, where $\phi_m(x)=\langle x|\phi_m\rangle$ and $\phi_l(x)=\langle x|\phi_l\rangle$.
Therefore, even in the case with complex spectrum, if the Hamiltonian is \textit{T}~symmetric [i.e.~condition \textit{(b)}], we have from Eqs.~(\ref{eq.t+})-(\ref{eq.t-}) 
$\langle x (+\infty)\rangle = \langle x (-\infty)\rangle$,
and hence Eq.~(\ref{eqdiag2}). For a Hamiltonian $H$ with \textit{PT} symmetry (without \textit{T} symmetry) we do not obtain Eq.~(\ref{eqdiag2}) because, using the properties above, we get $|\phi_{m}\rangle =$ $\mathcal{PT}|\phi_{l}\rangle$, $\phi_{m}(x)=\phi_{l}^*(-x)$ and hence $\langle x (+\infty)\rangle =$ $- \langle x (-\infty)\rangle$. 


Now, we assume condition~\textit{(c)} without requiring $H$ to be
\textit{T} or \textit{PT} symmetric. 
Therefore, for each disorder realization $H$, its parity-time counterpart $\tilde{H}={\mathcal{PT}} H({\mathcal{PT}})^{-1}$ is also a disorder realization of the same model. 
Then, inserting $(\mathcal{PT})^{-1}\mathcal{PT}$ in the expression for $\langle x(t)\rangle$ we find
\begin{eqnarray}
\langle x(t)\rangle_{ H}
&=&\frac{\langle \psi_0|\textrm{exp}(+i H^\dagger t)\, X\,\textrm{exp}(-i H t) |\psi_0\rangle}{\langle \psi_0 |\textrm{exp}(+i H^\dagger t)\, \textrm{exp}(-i H t)| \psi_0\rangle}\nonumber\\
%
%
%
&=&\frac{\bra{\psi_0}  
\textrm{exp}(-i\tilde{H}^\dagger t) (-X)\, \textrm{exp}(i\tilde{H} t)
\ket{\psi_0}}{\bra{\psi_0}  
\textrm{exp}(-i\tilde{H}^\dagger t) \,\textrm{exp}(i\tilde{H} t)
\ket{\psi_0}}\nonumber\\
&=& -\langle x(-t)\rangle_{\tilde{H}},
\end{eqnarray}
where we have used condition~\textit{(d)}.
From condition~\textit{(c)}, we conclude that $\overline{\langle x(t)\rangle} = -\overline{\langle x(-t)\rangle}$ 
and, in particular, 
\begin{eqnarray}
\overline{\langle x(+\infty)\rangle} &=& -\overline{\langle x(-\infty)\rangle}.\label{eqder}
\end{eqnarray}
From Eqs.~(\ref{eqdiag2}) and (\ref{eqder}) 
we have
\begin{eqnarray}
\overline{\langle x(+\infty)\rangle}=0\label{vanish}
\end{eqnarray}
which guarantees that the boomerang effect occurs.

Note that Eq.~(\ref{eqder}) remains valid if the ensemble of disorder realizations is composed only by the two elements $H$ and $\tilde{H}={\mathcal{PT}} H({\mathcal{PT}})^{-1}$. If $H$ is \textit{PT} symmetric then $\tilde{H}=H$, condition~\textit{(c)} is trivially met and Eq.~(\ref{eqder}) is valid for each disorder realization. If the model has \textit{T} symmetry then $\tilde{H}=\mathcal{P}H\mathcal{P}^{-1}$ and condition~\textit{(c)} is equivalent to $\mathcal{P}\{H\}\mathcal{P}^{-1}=\{H\}$. If the \textit{T}-symmetric $H$ has complex eigenenergies, once Eq.~(\ref{eqdiag2}) can be obtained without averaging over many disorder realizations, a boomerang effect is expected for the single pair of realizations $H$ and $\tilde{H}=\mathcal{P}H\mathcal{P}^{-1}$ in the sense that one should obtain Eq.~(\ref{vanish}) using these two realizations.

We note that in the case of complex spectrum with localized states the boomerang-like behavior in principle might depend on the system size. This is so because, unlike the case with real spectrum, the wave packet is allowed to move to positions very far from its origin. If the eigenstate $|\phi_l\rangle$ with largest value for the imaginary part of its eigenenergy is localized far from the origin, it has an exponentially small overlap $c_l$ with $|\psi_0\rangle$. This means that the wave function will take a longer time to become $|\psi(t)\rangle\approx |\phi_l\rangle$. The more one increases the system size, the more the wave function may take to reach its final state. Therefore we cannot guarantee a boomerang-like behavior at finite times if in the case of complex spectrum we take the limit of large systems. Indeed, as will be discussed in the next section, our results suggest that the boomerang-like behavior can only be observed for finite chains.

Interestingly, a similar behavior can be observed in delocalized systems under certain circumstances. If the system size is finite and the spectrum is real one can obtain the diagonal ensemble for the time average of the center of mass over a large enough time interval. 
However, the time scale at which the boomerang-like behavior can be observed also
grows with the system size and eventually this behavior disappears in the limit of large systems.
The QBE, on the other side, differs from the phenomenon described above in the sense that it does not require temporal average and it is present no matter how large the system is. 

As a final comment, we emphasize that AL, the symmetry of the ensemble $\{H\}$ and the symmetry of the initial state are crucial for the observation of the QBE, both in the Hermitian and in the non-Hermitian case that we investigate in this work. 
As shown above, Hermiticity is not a strict requirement for the boomerang effect. 
Moreover, 
if the Hamiltonian is \textit{T} symmetric a weak boomerang effect (which depends on the system size) can still appear in cases with complex eigenenergies. 
Furthermore, our analytical arguments are valid for any dimensions. 


\begin{table*}[t]
\caption{\textbf{Presence of the quantum boomerang effect as a function of the symmetry and the phase of the system}. For each symmetry of the Hamiltonian (\textit{T}, \textit{PT} or none of these) and for each phase of the system (Anderson localized with real spectrum, complex spectrum or delocalized phase) we show with \cmark~the presence or with \xmark~the absence of the QBE. We also show the models we used to check the emergence of the QBE.
In this table we assume that conditions \textit{(c)} and \textit{(d)} are valid. *In the AL phase with complex spectrum of the random hopping model we find that the boomerang-like effect depends on the system size.}\label{tablee}
\begin{tabular}{c|l|l|l}
    \hline \hline
     & AL with real spectrum & AL with complex spectrum & Extended phase\\ \hline \hline
    \textit{T} & ~\cmark~ $\begin{matrix} \textrm{HN model} \\ \textrm{Random hopping model} \end{matrix}$ & ~\cmark~  *Random hopping model & ~\xmark~ HN model\\ 
    \hline
   \textit{PT} &  ~\cmark~ \textit{PT}-AA model & ~\xmark~ Another \textit{PT}-AA model \cite{Longhi2019} & ~\xmark~ \textit{PT}-AA model\\ 
    \hline
    none & ~\cmark~ \textit{T}-broken random hopping & ~\xmark~ $\begin{matrix} \textit{T}\textrm{-broken random hopping} \\ \textrm{HN with complex potential}\\ \textrm{Anderson with gain and loss} \\ \textrm{Random gain and loss}  \end{matrix}$ & ~\xmark~ \textrm{HN with complex potential}  \\
    \hline \hline
\end{tabular}
\end{table*}

\section{Models and numerical results}\label{model}
In this section we shall investigate the QBE in several 1D non-Hermitian lattice models. All considered models can be written in the general form
\begin{eqnarray}\label{generaleq}
H=\sum_{j} \left[-J_{j}^R c_{j+1}^\dagger c_j - J_{j}^L c_{j}^\dagger c_{j+1} +\epsilon_j c_j^\dagger c_j \right],
\end{eqnarray}
where 
$c_j^\dagger$ ($c_j$) creates (destroys) a particle on site $j$, $j=1$, $2$, $\cdots$, $N$. 
$J_{j}^R$ ($J_{j}^L$) characterizes the hopping of the particle to the right (left) site and $\epsilon_j$ is the onsite potential. 
The parameters $J_{j}^R$, $J_{j}^L$ and $\epsilon_j$ may be chosen to be real or complex and may lead to Hermitian or non-Hermitian models, models with \textit{T} symmetry, \textit{PT} symmetry or none of these symmetries. 
These parameters may be random, pseudo-random or homogeneous. 
The simplest case is the Hermitian Anderson model, realized with real hoppings $J_j^R=J_j^L=J$ and real onsite potential $\epsilon_j$ randomly sampled from a uniform
distribution over the interval $[-W/2,W/2]$. When $W>0$, all states are localized and the model presents the QBE~\cite{Prat2019Quantum,Tessieri2021Quantum}.



To investigate the QBE, we initialize the system in a Gaussian wave packet 
\begin{equation}
\psi_0(x_j)=\mathcal{N} \exp(-x_j^2/2\sigma^2+ik_0 x_j),
\end{equation}
where $\mathcal{N}$ is a normalization constant, $\sigma^2$ is the variance, $k_0$ is the initial momentum and $x_j$ is the position of site $j$ (for simplicity we consider a unitary lattice parameter $a=1$). This wave function satisfies condition~\textit{(d)} because $\mathcal{PT} |\psi_0 \rangle=+ |\psi_0 \rangle$.
Once a model is chosen (e.g. the Anderson model or HN model), we make $n_d$ disorder realizations and propagate the wave packet under the corresponding Hamiltonian for each of the realizations using a standard 4th-order Runge-Kutta method.
The calculations are done considering open boundary conditions, and typically the size $N$ of the system is large enough so that the wave function is negligibly small near the edges for all times considered in the propagation.
For non-Hermitian models, the norm of the wave functions might be not unitary after time evolution. Hence we normalize the wave functions $\psi(x_j,t)$ of each disorder realization at each time $t$. Finally, we take the average of $|\psi(x_j,t)|^2$ over all the disorder realizations in order to compute the center of mass $\overline{\langle x (t)\rangle }$, second moment $\overline{\langle x^2 (t)\rangle }$ and variance $\overline{\sigma(t)}^2={\overline{<x^2(t)>}-\overline{<x(t)>}^2}$~\cite{Prat2019Quantum}.

We briefly comment on the phenomenology that we observe in the numerical simulations. When conditions \textit{(a)-(d)} are met the QBE is present: the wave packet initially propagates balistically with momentum $k_0$ and $\overline{\langle x (t)\rangle }$ increases with $t$ up to a time $t_U$ where $\overline{\langle x (t)\rangle }$ makes a U-turn. Then $|\overline{\langle x (t)\rangle }|$ decreases with time and vanishes at the limit $t\to +\infty$. 
The dependence of the QBE on the symmetry of the Hamiltonian and on the phase of the system is shown schematically in table~\ref{tablee}, which also shows the models we used in our simulations.
In models without \textit{T} or \textit{PT} symmetry we find that, if the spectrum is complex, $\overline{\langle x (t)\rangle }$ typically presents a local minimum at $t_m>t_U$, and increases again for $t>t_m$. 
We show in the Appendix the evolution of the variance of the wavepacket and the failure of the dynamical relation, Eq.~(\ref{dynamical}), for non-Hermitian models.

\subsection{Models with \textit{T} symmetry}
In this section we analize two prototypical non-Hermitian \textit{T} symmetric models which display QBE.
We numerically illustrate this result focusing on the HN model with real onsite disorder and in a non-Hermitian random hopping model.

\subsubsection{Hatano-Nelson model with real onsite disorder}\label{Hatano-Nelson_model}
The standard HN model is a non-Hermitian generalization of the Anderson model~\cite{Anderson1958Absence} and is described by the Hamiltonian~(\ref{generaleq}) by taking real, non-symmetric hoppings $J_j^R=J+J_a$, $J_j^L=J-J_a$, where $J$ and $J_a$ are the symmetric and anti-symmetric components of the hopping amplitudes, respectively. In the standard HN model one also considers real onsite potential $\epsilon_j$ randomly sampled from a uniform
distribution over the interval $[-W/2,W/2]$.
In the case of periodic boundary conditions (PBC) this model presents a localization transition which coincides with a topological transition and a real-complex transition. The phase where all states are localized is a topologically trivial phase and all its eigenenergies are real. However, the phase where some of the eigenstates are extended possesses a nontrivial winding number and some eigenenergies are complex. The transition between these two phases may be achieved varying either $W/J$ or $J_a/J$~\cite{Gong2018}.
With open boundary conditions (OBC), which we employ in the present calculations, the real-complex transition and the AL transition do not coincide~\cite{Gong2018,Ueda2019,Li2021}.

In this model we denote by $\overline{( ... )_>}$ the average over disorder realizations of a given variable [such as $|\psi(x,t)|^2$, $\langle x(t)\rangle$ or $\langle x^2(t)\rangle$] in the case where we have chosen a \textit{positive} value for $J_a$. In other words, $\overline{( ... )_>}$ is the average over the ensemble $\{H\}_>$ of disorder realizations with fixed $J_a=J_0>0$. Similarly, $\overline{( ... )_<}$ is the average over
the ensemble $\{H\}_<$ of disorder realizations with a fixed \textit{negative} value for $J_a=-J_0$. 
For example, $\overline{\langle x \rangle_>}$ is the average of $\langle x \rangle$ over all the disorder realizations of the onsite potentials once we have chosen a value for $W$ and a \textit{positive} $J_a$, say $J_a/J=0.5$.

\begin{figure}[b]
\centering
\includegraphics[width=\columnwidth]{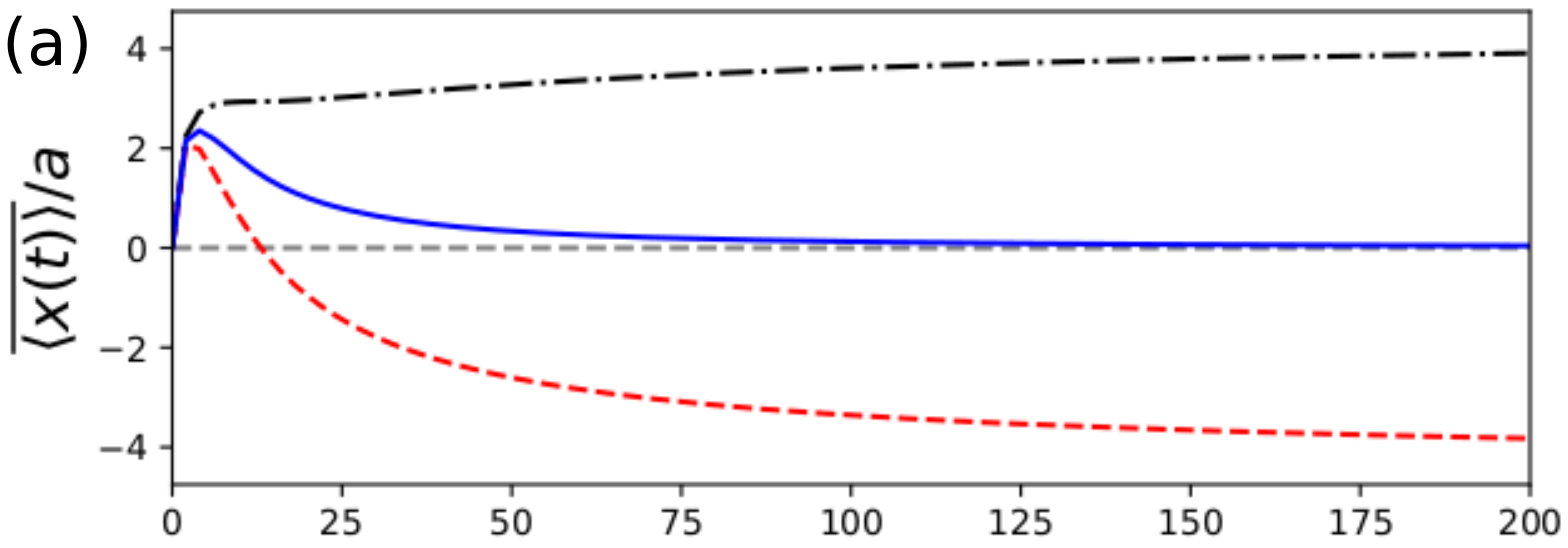}
\includegraphics[width=\columnwidth]{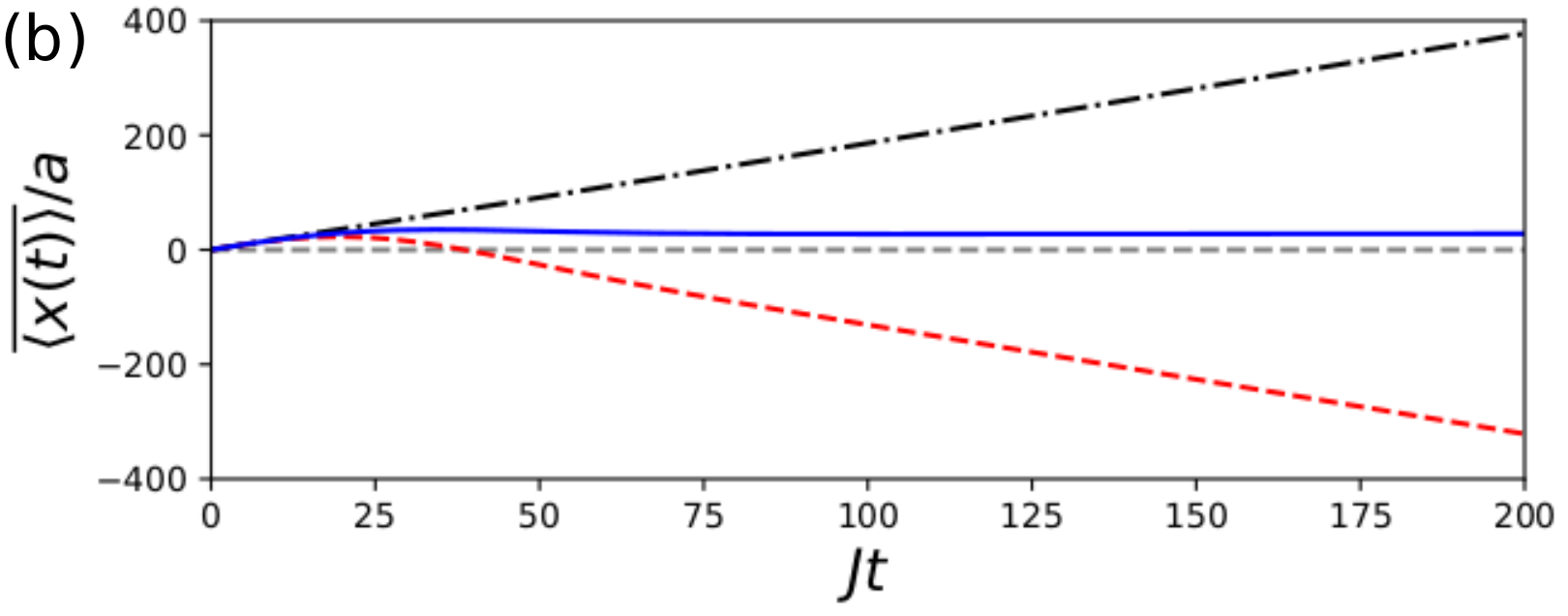}
\caption{\textbf{Quantum boomerang effect in the Hatano-Nelson model}. The curves show the disorder-averaged center of mass as a function of time for $|J_a|/J=0.01$ and (a) $W/J=3$, where all states are Anderson localized and (b) $W/J=0.5$, where there are extended states. We used $k_0a=1.4$, $\sigma/a=10$, $N=5\times 10^3$ sites and $n_d=10^6$ disorder realizations. Black dot-dashed (red dashed) lines were obtained with $J_a>0$ ($J_a<0$). The blue solid lines were obtained with the average of the two previous lines and show the presence of the QBE in (a) and its absence in (b). The gray dashed lines show the axis $x=0$. For both panels the spectrum is real.
}\label{boomerang_HNa}
\end{figure}

The HN model is \textit{T} symmetric.
The hopping terms of the Hamiltonian, which can be written in the form 
$H_0=\sum_{j} \big[-J^R c_{j+1}^\dagger c_j$ $- J^L c_{j}^\dagger c_{j+1} \big]$,
satisfy
$\mathcal{P}H_0\mathcal{P}^{-1}=\sum_{j} \left[-J^R c_{j}^\dagger c_{j+1} - J^L c_{j+1}^\dagger c_{j} \right]\neq H_0$.
This means that if we choose in the Hamiltonian $H$ the hopping to the right larger than the one to the left $J^R>J^L$, which is equivalent to $J_a>0$, we have that in $\mathcal{P}H\mathcal{P}^{-1}$ the hopping to the left is larger than the one to the right, and vice-versa.
Therefore $\mathcal{P}\{H\}_>\mathcal{P}^{-1}=\{H\}_<$.
%
Hence the analytical arguments of Sec.~\ref{analytical} do not predict the QBE in $\overline{\langle x(t) \rangle_>}$ or $\overline{\langle x(t) \rangle_<}$. 
Instead, we choose a value for $|J_a|$ and consider $\overline{( ... )}=\left[\overline{( ... )_>}+\overline{( ... )_<}\right]/2$ as the average over the disorder realizations in $\{H\}=\{H\}_> \cup  \{H\}_<$, which satisfy $\mathcal{P}\{H\}\mathcal{P}^{-1}=\{H\}$.
Then we can expect the QBE in $\overline{\langle x(t) \rangle}$ according to our analytical prediction. 
Notice that averaging over $\{H\}=\{H\}_> \cup  \{H\}_<$ is not equivalent to average over disorder realizations of the Anderson model, where $J^R=J^L=J$. The resulting DACM
$\overline{\langle x(t) \rangle}$ is different in these two models.

Using OBC we check numerically that in the localized regime (i.e.~$W>W_c$, where $W_c$ is the critical disorder) the spectrum is real and $\lim_{t\to+\infty} \overline{\langle x (t) \rangle_>}=-\lim_{t\to+\infty} \overline{\langle x (t) \rangle_<}\neq 0$. 
Therefore, for any initial momentum $k_0$ we find $\lim_{t\to+\infty} \overline{\langle x (t) \rangle}=0$ and the QBE is present in $\overline{\langle x (t) \rangle}$ but not in $\overline{\langle x (t) \rangle_\lessgtr}$, see Fig.~\ref{boomerang_HNa}(a). 
We note that using a small magnitude of asymmetry of the hoppings $|J_a|/J$ may already lead to a relatively large value of $\lim_{t\to+\infty} \overline{\langle x (t) \rangle_\lessgtr}/a$. 
%
Further increasing $|J_a|/J$ and keeping $W/J$ fixed causes $\lim_{t\to+\infty} \overline{\langle x (t) \rangle_\lessgtr}$ to be even larger. For large enough values of $|J_a|/J$ some of the eigenstates of the Hamiltonian will become extended. 
%
This transition can also be reached by fixing $|J_a|/J$ while decreasing $W/J$.
In this regime with extended eigenstates we observe a linear behavior for the DACM at large times, i.e.~for $1\ll Jt$ we have $\overline{\langle x (t) \rangle_\lessgtr}\sim bt$, with $b\lessgtr0$ [see Fig~\ref{boomerang_HNa}(b)].
In this phase we have $\lim_{t\to\infty}\overline{\langle x (t)\rangle}\neq 0$ and the QBE is not present. 

\begin{figure}[t]
\centering
\includegraphics[width=\columnwidth]{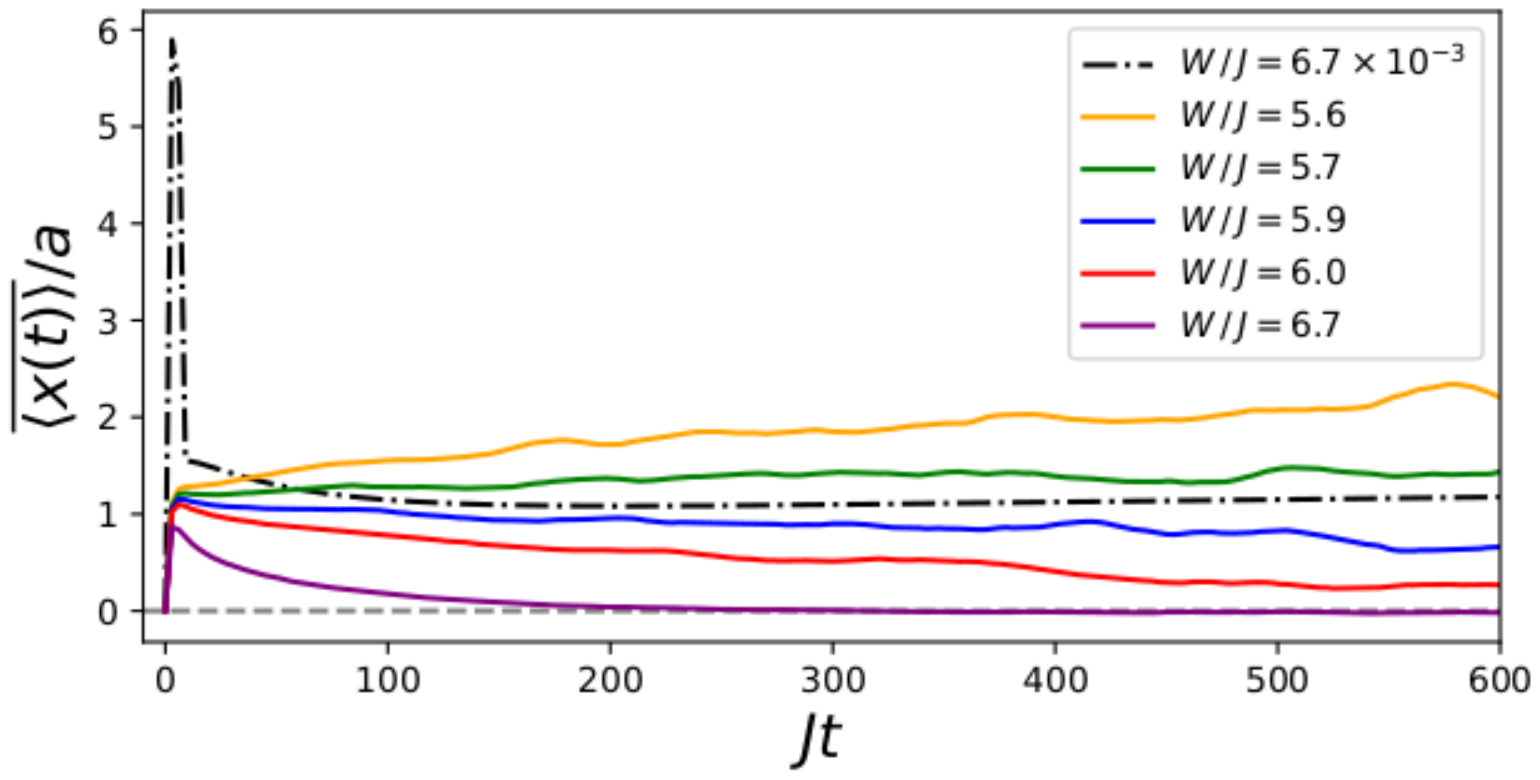}
\caption{\textbf{Probing localization transition using quantum boomerang.} Disorder-averaged center of mass $\overline{\langle x \rangle}$ as a function of time in the HN model for 
$|J_a|/J=1/3$ and several values of disorder $W/J$. 
Here we used $\sigma/a=10$, $k_0a=1.4$, $N=4\times 10^3$ sites and $n_d=10^6$ disorder realizations. 
The QBE is present for $W/J>W_c/J\approx 5.7$. For all lines with $W/J\geq 5.6$ the spectrum is real.
}\label{transitionUeda}
\end{figure}
In Fig.~\ref{transitionUeda} we investigate the extended-localized transition and its influence on the QBE. 
Here, to compare our results with the transition found in Ref.~\cite{Gong2018}, we use $|J_a|/J=1/3$.
For PBC the transition takes place at $W_c/J\approx 5.7$~\cite{Gong2018}. It can be clearly seen in Fig.~\ref{transitionUeda} that for $W<W_c$ we have $\lim_{t\to\infty}\overline{\langle x (t)\rangle}>0$ and the QBE is absent. 
For $W>W_c$ there is a U-turn and we have $\lim_{t\to\infty}\overline{\langle x (t)\rangle}=0$ in agreement with the QBE.
Importantly, as in the Hermitian case, the appearance of this effect can be used to find the critical disorder $W_c$ of the model~\cite{Prat2019Quantum}.

Here we have shown the presence of the QBE in the HN model when all the states are localized. An additional discussion on the numerical results in this model can be found in Appendix~\ref{app.A}. 
We also verify that in the localized regime the variance $\overline{\sigma(t)}$ saturates as $t\to +\infty$ 
and the dynamical relation, Eq.~(\ref{dynamical}), is a reasonable approximation only if the non-Hermiticity is small, i.e.~$|J_a|/J\ll 1$ (see Appendix~\ref{app.B}).
In the following we investigate another non-Hermitian model with \textit{T} symmetry. 

\subsubsection{Random hopping model}\label{Random hopping model}

Reference~\cite{Tessieri2021Quantum} considered a band random matrix model, which is a Hermitian, \textit{T}-symmetric model that presents the QBE. The authors considered random hopping amplitudes between the first $l$ neighbors. The model with random hoppings up to first neighbors can be obtained from Eq.~(\ref{generaleq}) by considering real parameters $J_j^R=J_{j}^L=J+h_j$, where $h_j$ are random numbers sampled from a uniform
distribution over $[-W_b/2,W_b/2]$. The onsite potentials $\epsilon_j$ are randomly sampled from a uniform
distribution over $[-W/2,W/2]$. It was shown in Ref.~\cite{Tessieri2021Quantum} that one has to consider $J\neq 0$ to be able to observe the QBE, otherwise $\overline{\langle x (t) \rangle}$ is negligibly small at all $t$. It was also shown that the amplitude of the boomerang $\overline{\langle x (t_U) \rangle}$ decreases when one increases $W_b$.

\begin{figure}[b]
\centering
\includegraphics[width=\columnwidth]{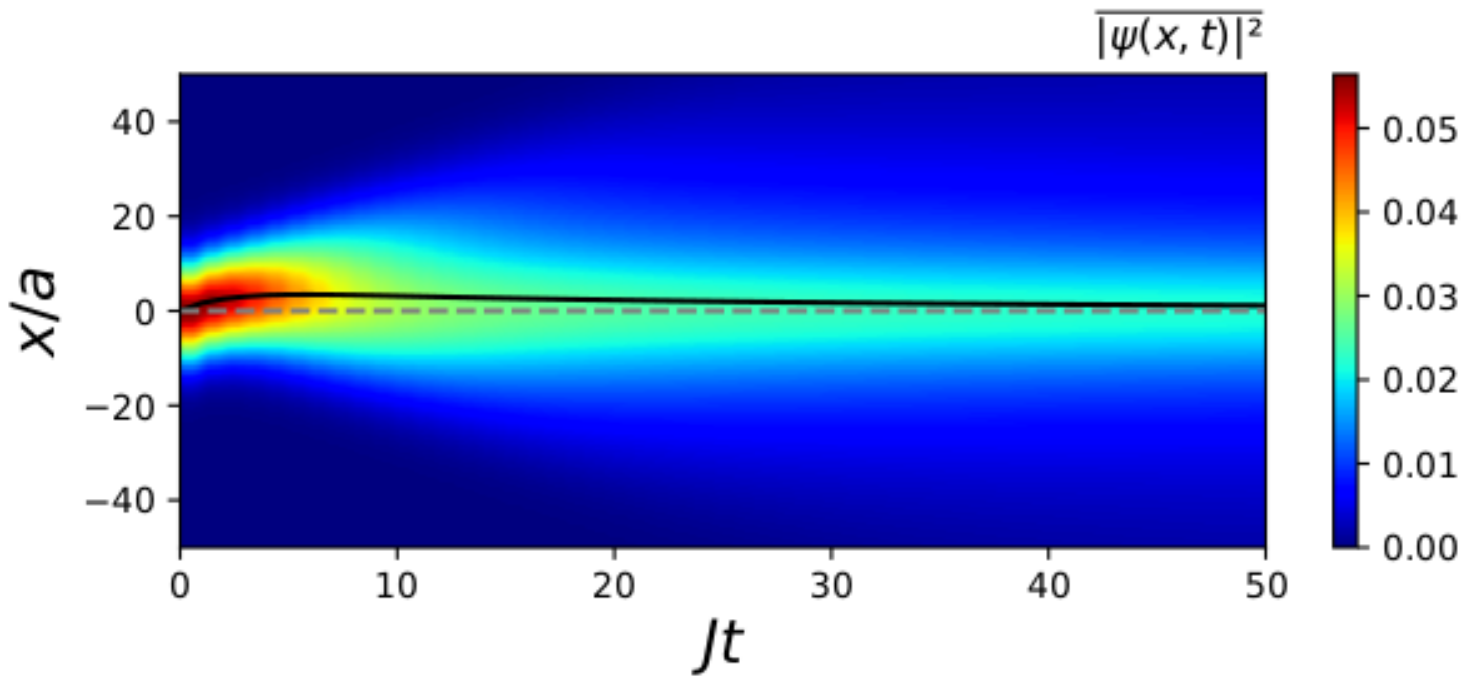}
\caption{\textbf{Boomerang effect in the \textit{T}-symmetric non-Hermitian random hopping model}. 
Using $W/J=2$ and $W_b/J=1$, which correspond to a phase with real eigenenergies, we show 
a density plot of the disorder-averaged probability distribution. Black solid line shows its center of mass as a function of time and gray dashed line shows the axis $x=0$. We used $\sigma/a=10$, $k_0a=1.4$, $N=2\times 10^3$ and $n_d=10^6$.}\label{randomhop}
\end{figure}
Here we consider a modification of that model that leads to a non-Hermitian, \textit{T}-symmetric random hopping model. This is done considering independent hoppings to the right and to the left, $J_j^R=J+h_j^R$, $J_{j}^L=J+h_j^L$, where $h_j^R$ and $h_j^L$ are uncorrelated random numbers sampled from a uniform distribution over $[-W_b/2,W_b/2]$. This model meets condition~\textit{(c)}.
Using $W/J=2$ and $W_b/J=1$, we find that the system is in a phase where the states are localized, all eigenenergies are real and the QBE is present (see Fig.~\ref{randomhop}). 
As expected, in this localized phase the variance $\overline{\sigma(t)}$ saturates as $t\to +\infty$. We also find that the dynamical relation breaks down in the non-Hermitian random hopping model (see Appendix~\ref{app.B}).

\begin{figure}[t]
\centering
\includegraphics[width=\columnwidth]{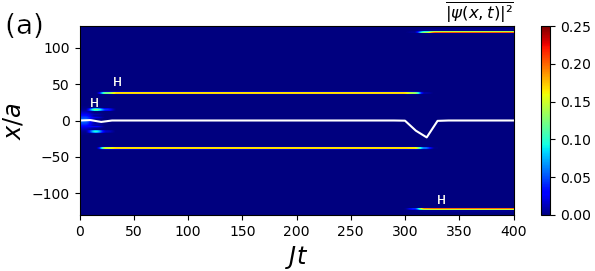}
\includegraphics[width=\columnwidth]{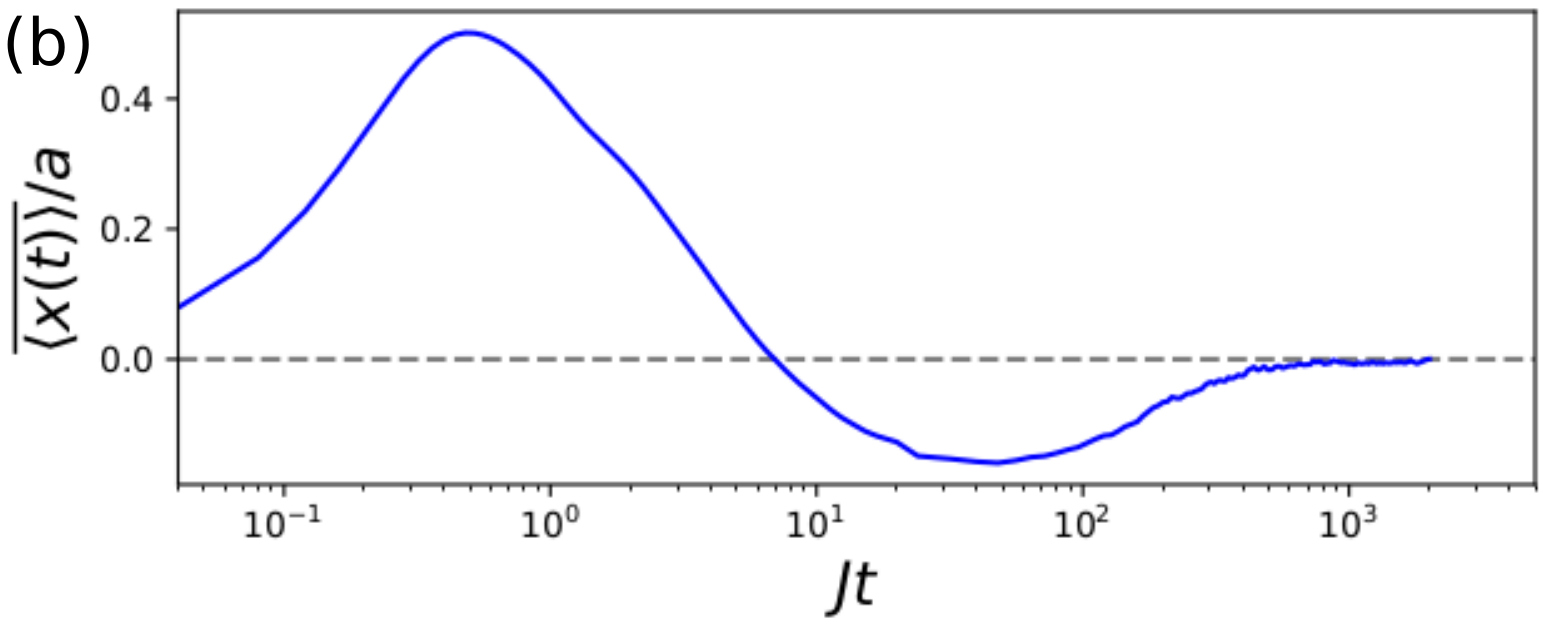}
\caption{\textbf{Boomerang in the non-Hermitian random hopping model with complex spectrum}. 
Using $W/J=10$, $W_b/J=3$, 
$\sigma/a=10$, $k_0a=\pi/2$ and $N=301$ we show in (a) the probability distribution averaged over two disorder realizations, $H$ and its parity-inversion $\mathcal{P}H\mathcal{P}^{-1}$. 
The wave functions are localized in space. The label H in the figure indicates the contribution of the probability distribution that comes from the disorder realization $H$, while the parity inverted contributions of the figure come from $\mathcal{P}H\mathcal{P}^{-1}$. At large times $t$ the wave functions reach the eigenstates whose eigenenergies have the largest values for their imaginary parts.
The disorder averaged center of mass $\overline{\langle x(t)\rangle}$ of these two realizations is given in white curve and vanishes at large times. (b) shows the disorder averaged center of mass of $n_d=10^6$ realizations and their corresponding parity inversions. Even in this case with complex spectrum the boomerang behavior is present.}\label{randomhop2}
\end{figure}

We verify that, increasing $W_b$, initially causes the amplitude of the center of mass at the U-turn $\overline{\langle x(t_U) \rangle}$ to decrease. This is in agreement with Ref.~\cite{Tessieri2021Quantum} in the Hermitian random hopping model, in which the localization length decreases as one increases $W_b$ or $W$. 
However, we find that in our non-Hermitian model increasing $W_b$ beyond a critical value $W_{b}/J=2$ causes the appearance of complex eigenenergies, while the states remain localized. 
The emergence of complex eigenenergies can be understood considering the simple case of a two-site model. In this case the Hamiltonian can be written as
\begin{equation}
    H=\left( 
    \begin{tabular}{c c}
        $\epsilon_1$ & $-J-h^L$ \\
         $-J-h^R$ & $\epsilon_2$
    \end{tabular}
    \right),
\end{equation}
and we find its corresponding eigenenergies to be $E_\pm=\frac{1}{2}\left[ 
    \epsilon_1+\epsilon_2\pm \sqrt{(\epsilon_1-\epsilon_2)^2+4(J+h^L)(J+h^R)}
    \right].$
When $W_b/J<2$ there cannot be complex eigenvalues. When $W_b/J>2$ there will be complex eigenvalues for some of the realizations of the random parameters $\epsilon_1,\epsilon_2,h^L,h^R$.

In our analytical demonstration of the boomerang effect we required all eigenenergies to be real, except in the case of \textit{T}-symmetric models [condition \textit{(b)}], such as the random hopping model we are studying here. Once the eigenstates are localized in this phase, a boomerang-like behavior is expected, even in the case of complex spectrum. 
In Fig.~\ref{randomhop2} we show a detailed investigation with evidence of the boomerang behavior. Panel~(a) illustrates that the wave function tends to be localized in states whose eigenenergies have large imaginary parts $\varepsilon''$. It also shows that the DACM of two disorder realizations $H$ and $\mathcal{P}H\mathcal{P}^{-1}$ is essentially zero, except for short intervals when the wave function is migrating to eigenstates with larger $\varepsilon''$. For large enough times, the DACM will remain equal to zero. In panel~(b) we see in the case of many disorder realizations that the DACM tends to vanish for large $t$, indicating the presence of the boomerang behavior even in the case of complex spectrum, once the model is \textit{T} symmetric.
As mentioned in Sec.~\ref{analytical}, this behavior was expected for the case of complex spectrum with finite system size and long enough times. In Fig.~\ref{randomhop2}, in order to see this boomerang-like behavior, we consider a smaller lattice and only observe the boomerang at long times. As we increase the size of the chain we observe that the DACM tends to saturate at a finite position for the time scale we can numerically compute, being expected to vanish only for much larger times. This indicates that the boomerang behavior disappears in the limit of large system sizes, contrary to the conventional QBE that takes place in the case with real spectrum.

\subsection{Models with \textit{PT} symmetry}
In this subsection we shall investigate the presence of the QBE in a non-Hermitian, \textit{PT} symmetric generalization of the Aubry-Andr\'e (AA) model. To that aim we first demonstrate the QBE in the Hermitian AA model. Next we show results for its non-Hermitian generalization.

\subsubsection{Hermitian Aubry-Andr\'e model}

In Ref.~\cite{Tessieri2021Quantum} the authors investigated the QBE in a Hermitian model with real hoppings $J_j^R=J_j^L$ $=J$ and pseudorandom onsite potentials of the form
\begin{equation}
    \epsilon_j=W \cos{\left( \pi \sqrt{5} j^\gamma \right)},\label{eq.pseudorandom}
\end{equation}
where $j>0$, $\gamma>0$ and $W>0$ is the strength of the disorder. The localization of the eigenstates and the presence of QBE depends on the parameter $\gamma$. If $\gamma\geq 2$ all states are localized and the QBE is present in a very similar way of the Anderson model.
For $1<\gamma<2$ the state at the band center is delocalized, while the other states are localized with a longer localization length. In this case the QBE is still present, but its height and width depend on the position of the chain where the center of the wave packed is initially located. This may be due to the fact that in this regime the onsite potential has a slowly varying period for large values of the site index $j$. For $\gamma=1$ one has the AA model, which possesses a mobility edge at $W_c/J=2$. For $W<W_c$ ($W>W_c$) all eigenstates of the model are extended (localized). The presence of the QBE was not reported in this model. For $0<\gamma<1$ there are extended states and hence the QBE is not expected~\cite{Tessieri2021Quantum}.


%
\begin{figure}[tb]
\centering 
\includegraphics[width=\columnwidth]{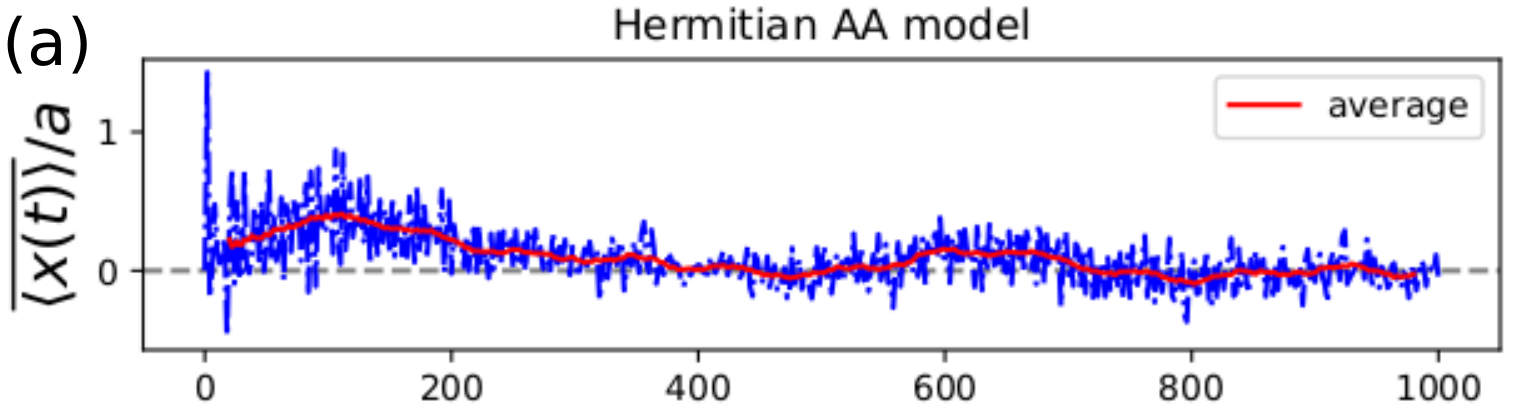}\vspace{0.2cm}
\includegraphics[width=\columnwidth]{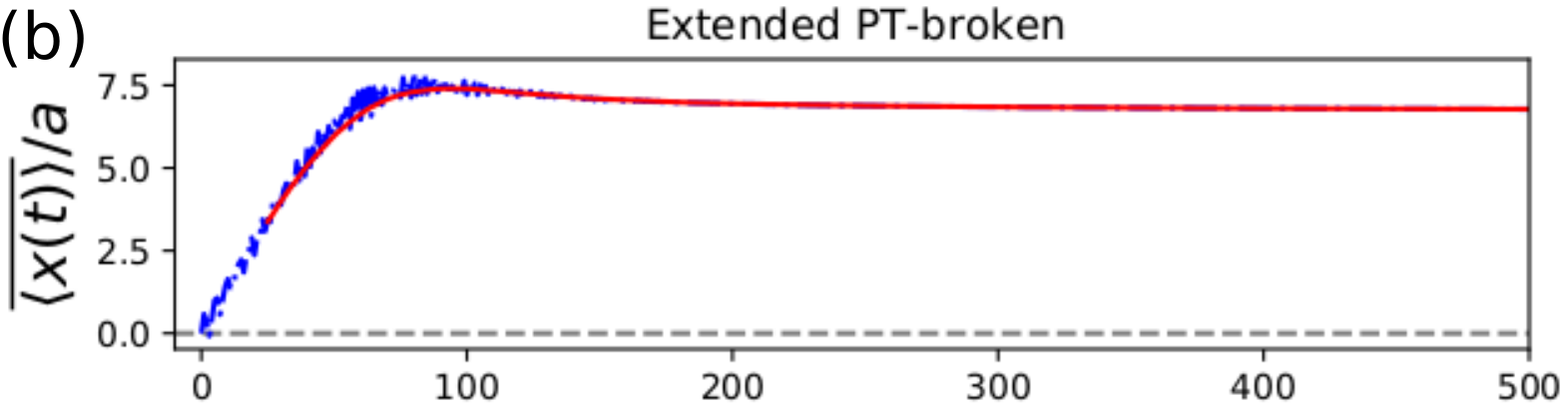}\vspace{0.2cm}
\includegraphics[width=\columnwidth]{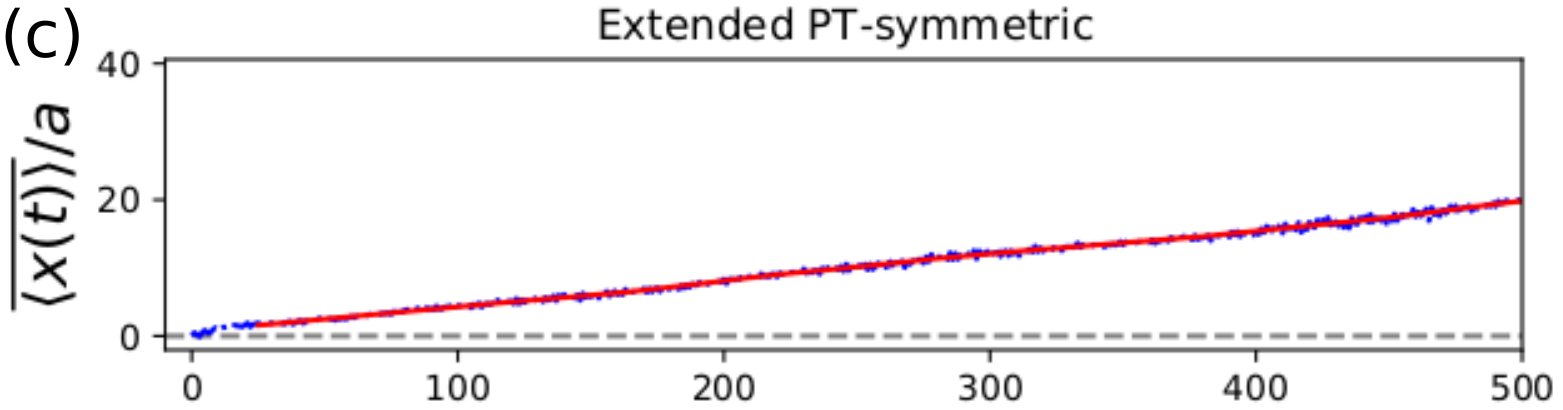}\vspace{0.2cm}
\includegraphics[width=\columnwidth]{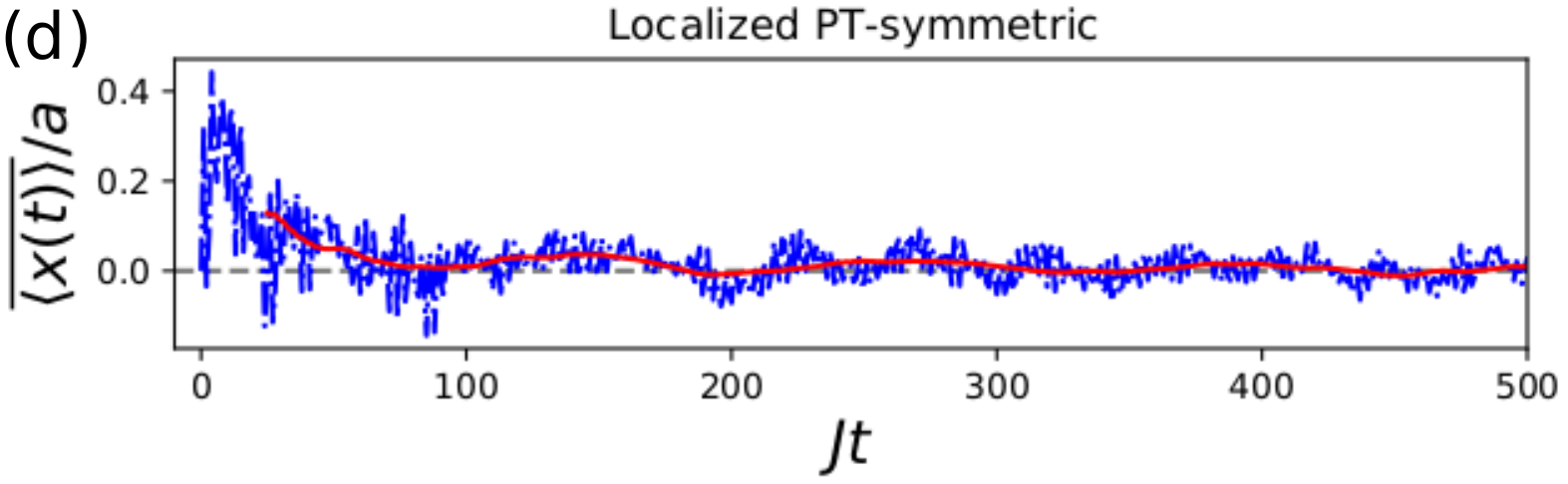}
\caption{\textbf{Boomerang effect in \textit{PT}-symmetric models}. The blue dot-dashed lines show $\overline{\langle x(t)\rangle}$. In (a) we consider the Hermitian Aubry-Andr\'e model and we used  $W/J=2.1$, $k_0a=1.4$, $\sigma/a=1$, $n_d=10^5$ and $N=10^3$. The red solid line shows the average of $\overline{\langle x(t)\rangle}$ over the interval $[t-20/J,t+20/J]$. 
In panels (b)-(d) we consider the non-Hermitian \textit{PT}-symmetric Aubry-Andr\'e model.
Here we used  
$\gamma_0/J=\tan{(\pi/3)}\approx1.73$, $k_0a=\pi/2$, $\sigma/a=1$, $n_d=10^5$, $M=2584$ and $N=4181$. These panels were obtained with: (b) $W/J=1.6$ and corresponds to the \textit{PT} broken phase with extended states; (c) $W/J=1.8$ and corresponds to the \textit{PT} symmetric phase with extended states; (d) $W/J=2.2$ and corresponds to the \textit{PT} symmetric phase with localized states, which presents the QBE. 
The red lines in (b)-(d) show the average of $\overline{\langle x(t)\rangle}$ over the interval $[t-12.5/J,t+12.5/J]$.
}\label{AA_model}
\end{figure}
Here we report the presence of the QBE in the Hermitian AA model.
To mimic the average over disorder realizations, we consider $n_d$ chains, obtained from a shift in the onsite potential $\epsilon_j$. The onsite potential for the $i$th chain is of the form $\epsilon_j^{(i)}=W\cos{\left[ \pi\sqrt{5}(j+i-1)  \right]}$. Figure~\ref{AA_model}(a) shows $\overline{\langle x(t)\rangle}$ in the Hermitian AA model with $W/J=2.1$ $>W_c/J$ and $n_d=10^5$ realizations. 
We observe an initial peak $\overline{\langle x(t_U)\rangle}\approx 1.4\, a$ followed by oscillations about an average (in red solid line) that goes to zero as $t\to+\infty$. These fluctuations in the QBE in the AA model are similar to the fluctuations in the QBE in the quantum kicked rotor, both in theory~\cite{Tessieri2021Quantum} and experiment~\cite{Sajjad2021Observation}. In both models, AA and quantum kicked rotor, one has psudorandom potentials. 
In the case of the AA model, the potential is of the form $\epsilon_j\sim \cos{(\beta j + \varphi)}$, while in the case of the kicked rotor it is of the form $\epsilon_p\sim\tan{(\beta p^2 + \varphi)}$ in momentum space~\cite{Fishman1982,Tessieri2021Quantum}. One possible cause for the oscillations in both models is the pseudorandom nature of the onsite potentials. 

In the AA model the variance saturates as $t\to+\infty$. On the one side, the standard AA model is Hermitian and the dynamical relation is expected to hold. On the other side, the large amount of fluctuations with time in this model could spoil the computation of the time derivative in Eq.~(\ref{dynamical}). We find however that the dynamical relation works reasonably well in this model (see Appendix~\ref{app.B}). Using $W<W_c$ we do not find the QBE in the AA model due to the presence of extended states in this regime. 

\subsubsection{Non-Hermitian Aubry-Andr\'e model with \textit{PT} symmetry}

Now we investigate the QBE in a non-Hermitian, \textit{PT} symmetric generalization of the AA model. 
One such model, discussed in Ref.~\cite{Scriffer2021Anderson}, may be obtained from Eq.~(\ref{generaleq}) choosing complex and asymmetric hoppings $J_j^R=J_{j-1}^L=J_j=J+i\gamma_0 \sin{(2\pi \beta j +\varphi)}\neq J_{j+1}^*$. 
The onsite potentials are real, $\epsilon_j=2W\cos{(2\pi \beta j +\varphi)}$. Here, $\gamma_0$ controls the non-Hermiticity, $W$ is the quasidisorder strength and $\beta=M/N$, where $M$ and $N$ are two adjacent Fibonacci numbers. One recovers the Hermitian AA model when $\gamma_0=0$.
%
%
It was proved that this Hamiltonian is \textit{PT} symmetric when $\varphi=m\pi/N$ if $m$ is odd (integer) and $N$ is even (odd).
It was also shown that depending on the values of $J/W$ and $\gamma_0/W$ this model may lead to three distinct phases: \textit{PT} broken phase, i.e.~with complex eigenenergies, with extended states (for $\gamma_0/W>1$); \textit{PT} symmetric phase, i.e.~with real eigenenergies, with extended states [for $\gamma_0/W<1$ and $(\gamma_0/W)^2+(J/W)^2>1$]; and \textit{PT} symmetric phase with localized states [for $(\gamma_0/W)^2+(J/W)^2$ $<1$]~\cite{Scriffer2021Anderson}.

In Figs.~\ref{AA_model}(b)-(d) we investigate the behavior of the center of mass $\overline{\langle x(t)\rangle}$ in the three phases of the model. 
It is interesting to note that each of the three phases presents a different behavior for $\overline{\langle x(t)\rangle}$, and the observation of the DACM could be used to find the transition between any of these phases. 
As expected, the QBE does not appear in the phases with extended states. However, the QBE does appear in the \textit{PT} symmetric phase with localized states. It displays fluctuations similiar to the ones that appear in the Hermitian AA model, probably due to the pseudorandom nature of the onsite potentials. 
The variance also presents a different behavior in each of the three phases (see Appendix~\ref{app.B}). We find that the dynamical relation breaks down in the phase that presents the QBE.


\subsection{Models without \textit{T} or \textit{PT} symmetry}
In the following we illustrate several non-Hermitian models without \textit{T} or \textit{PT} symmetry and show that the QBE appears only if conditions~\textit{(a)-(d)} are met.

\subsubsection{\textit{T}-broken non-Hermitian random hopping model}\label{TRandom}

\begin{figure}[b]
\centering 
\includegraphics[width=\columnwidth]{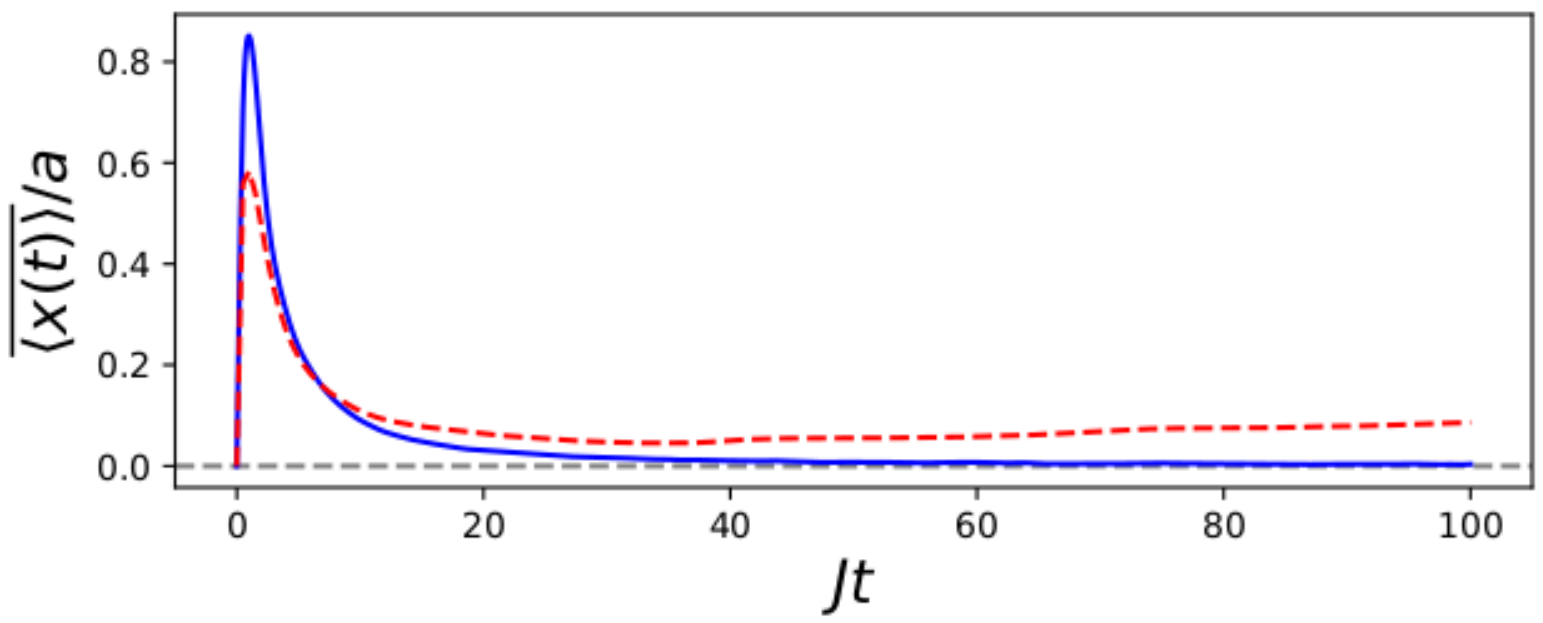}
\caption{\textbf{\textit{T}-broken random hopping model.} We compute  $\overline{\langle x(t) \rangle}$ using $W/J=6$, $k_0a=1.4$, $\sigma/a = 10$. In the blue solid line we use $W_b/J=1$, $N=2\times 10^2$ and  $n_d= 10^6$, the spectrum is real and the model presents the QBE. In the red dashed line we use $W_b/J=4$, $N=5\times 10^2$ and  $n_d= 10^7$, the spectrum is complex and the QBE is absent.}\label{Tbrokenrandomh}
\end{figure}

Here we consider a modification of the random hopping model of Sec.~\ref{Random hopping model} that leads to a non-Hermitian, \textit{T}-broken random hopping model. This is done considering in Eq.~(\ref{generaleq}) independent hoppings to the right and to the left, $J_j^R=(J+h_j^R)\textrm{e}^{i\phi_j}$, $J_j^L=(J+h_j^L)\textrm{e}^{-i\phi_j}$, where $h_j^R$ and $h_j^L$ are uncorrelated random numbers sampled from a uniform distribution over $[-W_b/2,W_b/2]$ and $\phi_j$ are sampled uniformly from the interval $[0,\Phi]$, $\Phi>0$.
Time-reversal symmetry is broken due to the random phases $\phi_j$ and the model is non-Hermitian because $h_j^R\neq h_j^L$ for $W_b>0$. The onsite potentials $\epsilon_j$ are randomly sampled from a uniform distribution over $[-W/2,W/2]$. 
We find that for $\Phi=2\pi$ the DACM is negligibly small, in agreement with the Hermitian case with $J=0$~\cite{Tessieri2021Quantum}. Therefore we use here $\Phi=1$, once the DACM is finite in this case.
We verify that the spectrum is real for $W_b/J<2$ and complex for $W_b/J>2$, exactly as was found in the \textit{T}-symmetric case of Sec.~\ref{Random hopping model}. The QBE is present in the case of real spectrum because conditions~\textit{(a)-(d)} are met (see Fig.~\ref{Tbrokenrandomh}). In this case the variance $\overline{\sigma(t)}$ saturates as $t\to+\infty$. For complex spectrum the QBE is absent. This shows the importance of condition \textit{(b)}. 
In the case of complex spectrum, $\overline{\langle x(t) \rangle}$ makes the standard U-turn and after that it presents a local minimum at $t=t_m>t_U$, after which it increases again. This local minimum after the U-turn, which has not been reported before in the literature about the QBE, also appears in other non-Hermitian models with complex spectrum and without \textit{T} and \textit{PT} symmetry, as we discuss in the following.

\subsubsection{Hatano-Nelson model with complex potential}

\begin{figure}[b]
\centering 
\includegraphics[width=\columnwidth]{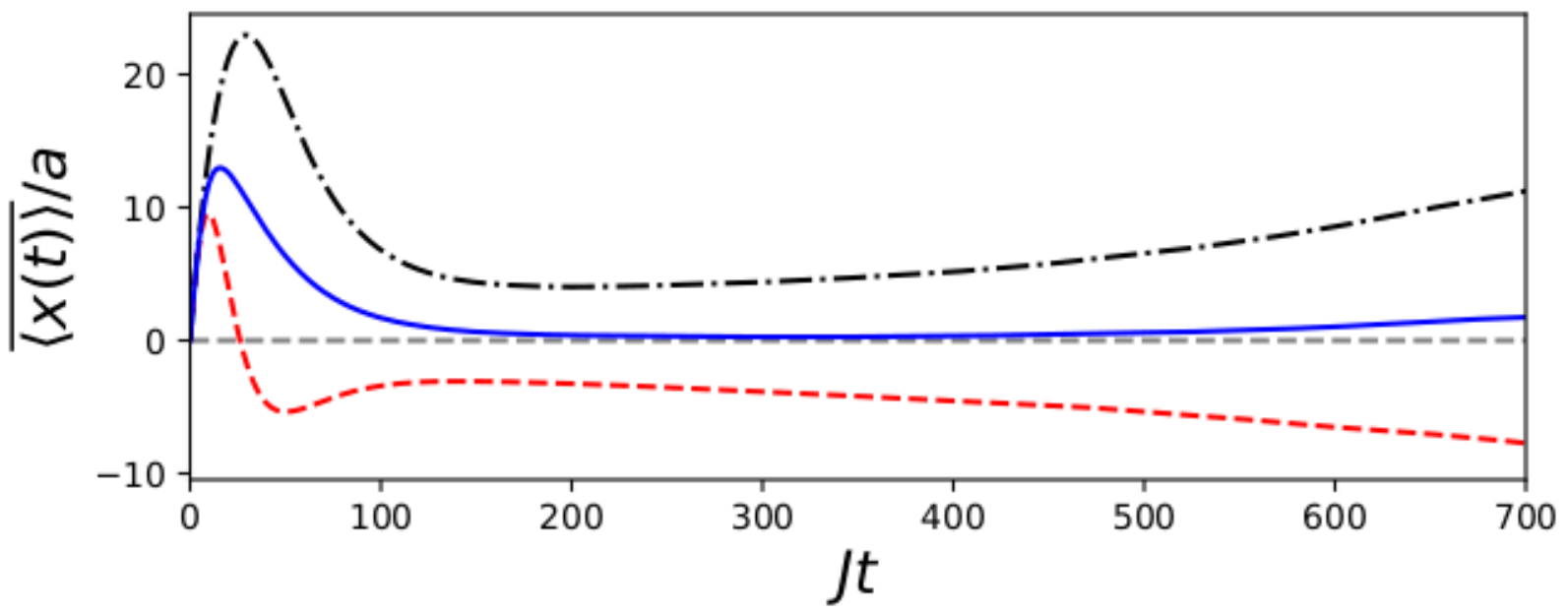}
\caption{\textbf{Hatano-Nelson model with complex potential.} We compute  $\overline{\langle x(t) \rangle}_>$ (black dot-dashed), $\overline{\langle x(t) \rangle}_<$ (red dashed) and $\overline{\langle x(t) \rangle}$ (blue solid line) using $|J_a|/J=0.01$, $W/J=0.5$, $k_0a=1.4$, $\sigma/a = 10$, $N=2\times 10^3$ and  $n_d=5\times 10^5$. The model does not present the QBE.}\label{boomerang_HNcomplex}
\end{figure}

In the standard HN model, it was considered in Eq.~(\ref{generaleq}) non-symmetric hoppings $J_j^R=J+J_a$, $J_j^L=J-J_a$, while the onsite potentials $\epsilon_j$ were chosen to be  random real numbers. In order to break \textit{T} symmetry in that model, here we consider complex onsite potentials     $\epsilon_j=|\epsilon_j|\textrm{e}^{i\phi_j}$, where $|\epsilon_j|\in [0,W]$ and $\phi_j\in [0,2\pi]$ are random numbers with uniform distribution.
This model, similarly to the standard HN model, possesses a localization transition. However, in contrast with the standard HN model, the present model with complex onsite potential has complex eigenenergies even in the localized regime, both for OBC and PBC~\cite{Gong2018}.
In this model we consider the ensemble of realizations given by $\{H\}=\{H\}_>\cup \{H\}_<$, as was done in the case of the standard HN model, to ensure that condition \textit{(c)} is met.
In Fig.~\ref{boomerang_HNcomplex} we show the DACM using parameters in the localized regime and verify that it does not present the QBE once condition \textit{(b)} is not met. 
Instead, the DACM presents a local minimum at $t=t_m>t_U$.

\subsubsection{Anderson model with alternating gain and loss}

\begin{figure}[b]
\centering 
\includegraphics[width=\columnwidth]{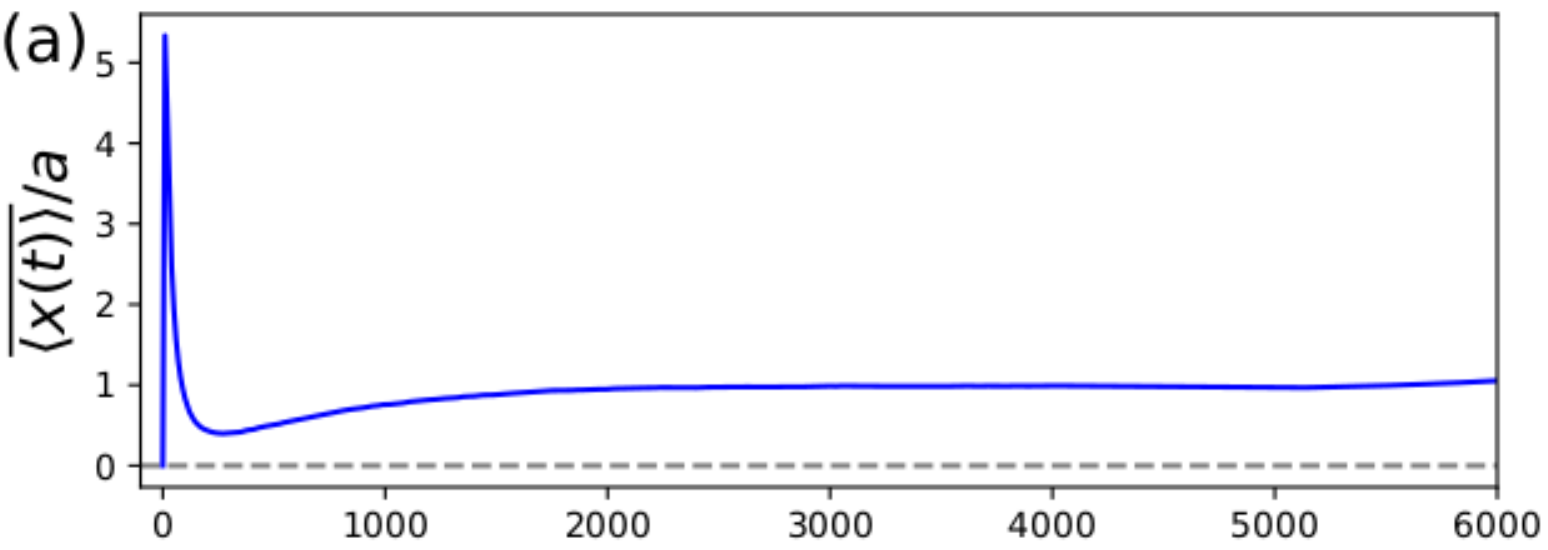}
\includegraphics[width=\columnwidth]{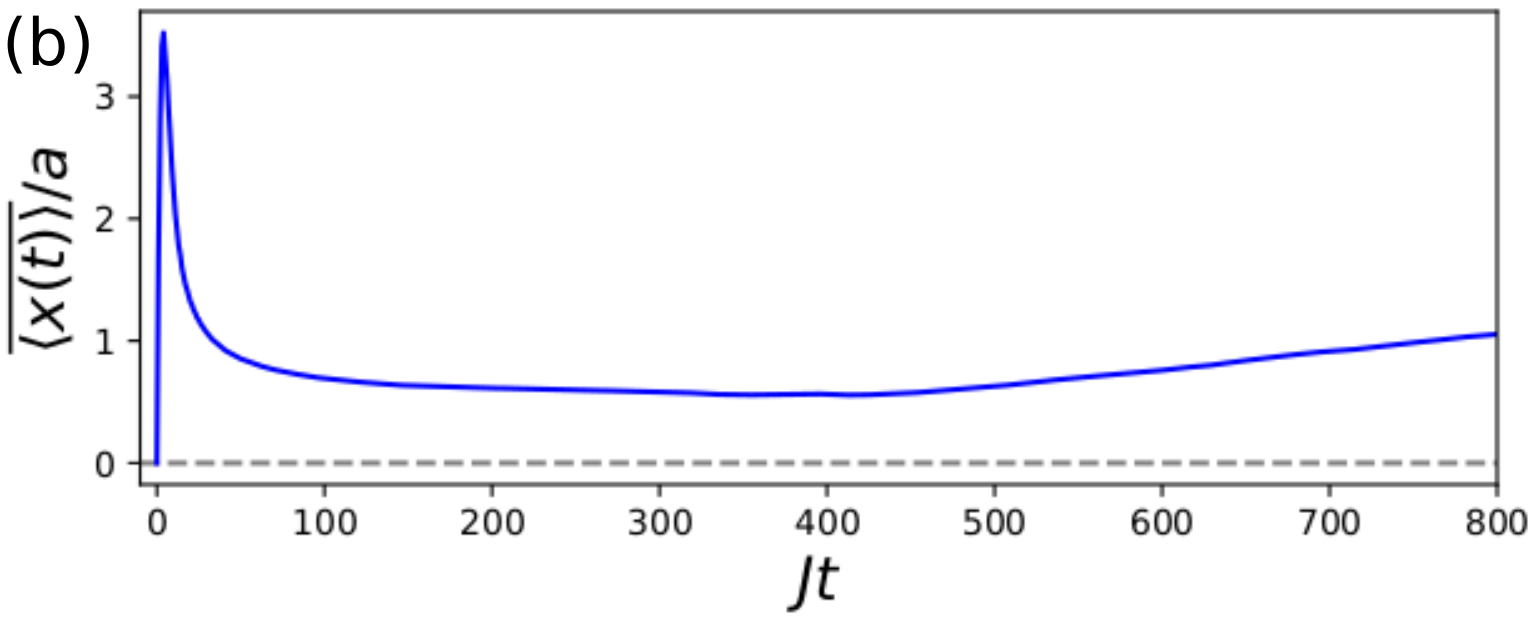}
\caption{\textbf{Center of mass in other models without \textit{T} or \textit{PT} symmetry.} 
(a) Anderson model with alternating gain and loss. Here we used $\gamma/J=0.01$, $W/J=2$, $k_0a=1.4$, $\sigma/a=10$, $n_d=10^6$ and $N=4\times 10^3$. 
(b) Random gain and loss model using $W/J=2$, $k_0a=1.4$, $\sigma/a=10$, $n_d=10^7$ and $N=2\times 10^3$. None of the models present the QBE. In both of them the center of mass presents a local minimum after the U-turn.} \label{anderson_gl}
\end{figure}

Now we consider the standard Anderson model \cite{Tessieri2021Quantum} with the addition of a gain and loss parameter ($\pm i \gamma$), which alternates from site to site.
This Anderson model with alternating gain and loss is obtained from the Hamiltonian~(\ref{generaleq}) by choosing real hoppings $J_j^R=J_j^L=J$ and complex onsite potential $\epsilon_j=\textrm{Re}(\epsilon_j)-i\gamma(-1)^j$, where $\gamma$ is a real parameter that controls the non-Hermiticity of the model and breaks \textit{T} symmetry. 
The disorder is encoded in the real part of the onsite potential, $\textrm{Re}(\epsilon_j)$, which has a uniform distribution over the interval $[-W/2,W/2]$~\cite{Hamazaki2019Non-Hermitian}.
%


The model has complex spectrum and condition \textit{(b)} is not met. Therefore we found that the QBE is not present, see Fig.~\ref{anderson_gl}(a). 
As one decreases the value of $\gamma$, the asymptotic value of $\overline{\langle x (t)\rangle}$ at long times decreases and the QBE is recovered when $\gamma=0$. 
%
%
For $\gamma\neq 0$ the DACM presents a U-turn and after that $\overline{\langle x (t)\rangle}$ acquires a local minimum value after which it slowly grows. 

\subsubsection{Random gain and loss model}
We consider here one more model without \textit{T} or \textit{PT} symmetry. In Eq.~(\ref{generaleq}) we choose real hoppings $J_j^R=J_j^L=J$ and purely imaginary onsite potentials $\epsilon_j=-ih_j$, where $h_j$ are random numbers sampled from a uniform distribution over $[-W/2,W/2]$. Figure~\ref{anderson_gl}(b) shows that the QBE is not present in this random gain and loss model, which also presents a local minimum in $\overline{\langle x(t) \rangle}$ at $t_m>t_U$. 
%


\section{Discussion and conclusion}\label{conclusion}

The QBE has been predicted to take place in several Hermitian models with Anderson localization and very recently has been experimentally observed in the quantum kicked rotor model. QBE predicts that the disorder-averaged center of mass of a particle launched with an initial velocity will initially move ballistically, make a U-turn, return to the origin and stop there.

In this manuscript we investigated the emergence of the quantum boomerang effect in non-Hermitian systems and clarified the importance of symmetries in the Hamiltonian and in the initial state. We have shown analytically that sufficient conditions to observe a boomerang effect in non-Hermitian systems are: \textit{(a)} Anderson localization, \textit{(b)}~reality of the eigenenergies or \textit{T} symmetry in the case of complex eigenenergies, \textit{(c)} \textit{PT} invariance of the ensemble of disorder realizations and \textit{(d)}
the initial wave function be an eigenstate of the parity-time operator.
Our arguments are valid for any dimension.

We confirm our analytical results through a careful numerical investigation of the dynamics in several non-Hermitian models. To study \textit{T}-symmetric models we considered the Hatano-Nelson model and a non-Hermitian random hopping model. As expected, we confirmed the boomerang effect whenever the eigenstates are localized and all eigenenergies are real. 
We also verify the boomerang effect when the eigenenergies become complex. However, in this case the boomerang behavior only appears if the system size is finite.
%
We investigated a non-Hermitian \textit{PT}-symmetric Aubry-Andr\'e model. 
The dynamics of the disorder-averaged center of mass and of the variance have a different behavior in each of the three possible phases of the model: extended \textit{PT}-broken, extended \textit{PT}-symmetric and localized \textit{PT}-symmetric phase. In this last case the boomerang effect is present, confirming the analytical prediction. 
We looked for the presence of the QBE in another non-Hermitian \textit{PT}-symmetric generalization of the Aubry-Andr\'e model, which is known to possess a phase where all states are localized but the eigenenergies are complex (i.e.~\textit{PT}-broken phase)~\cite{Longhi2019}. In this case the disorder-averaged center of mass does not return to the origin and the boomerang is absent. This suggests that our analytical arguments cannot be extended to guarantee the QBE in the case of complex eigenenergies in \textit{PT}-symmetric models.
In addition we investigated a \textit{T}-broken non-Hermitian random hopping model and found the boomerang in the phase with real spectrum once conditions \textit{(a)}-\textit{(d)} are met.
This model is more general than all other models presented in the literature about the QBE in the sense that it simultaneously breaks Hermiticity, \textit{T} symmetry, \textit{P} symmetry and \textit{PT} symmetry.
In any of the models with real spectrum, when the quantum boomerang effect is present we find that the variance saturates with time, as expected. However, the dynamical relation, which relates the center of mass with the time derivative of the second moment, breaks down as we increase non-Hermiticity in the models.

Investigating several non-Hermitian models without \textit{T} or \textit{PT} symmetry we show that the quantum boomerang effect is absent in the case of complex spectrum. 
We find that all of these models present a peculiar behavior: a local minimum of the disorder-averaged center of mass after the U-turn. This means that the center of mass initially departs from the origin, makes a U-turn toward the origin and, after some time, makes another U-turn moving away from the origin.

We comment on the experimental implementation of some non-Hermitian models discussed in the manuscript. 
Asymmetric hopping amplitudes were proposed for cold atoms in optical lattices in \cite{Gong2018}. 
Modeling the dynamics via a Lindblad master equation $\dot{\rho_t}=-i\left[H,\rho_t\right]$ $+\sum_j\mathcal{D}\left[L_j\right]\rho_t$, where $\mathcal{D}\left[L\right]\rho \equiv$ $L\rho L^{\dagger} - \left\{L^{\dagger}L,\rho \right\}/2$, one obtains an effective non-Hermitian Hamiltonian, $H_{\textrm{eff}}=H-\frac{i}{2}\sum_j L^{\dagger}_{j} L_{j}$~\cite{Lindblad1976} either under postselection or through loss processes in coherent condensates.
The effective non-Hermitian Hamiltonian for asymmetric hopping, $H_{\textrm{eff}} = \sum_j \left(J_R c^{\dagger}_{j+1}c_j +
J_L c^{\dagger}_{j}c_{j+1}
\right) -i\kappa \sum_j c_{j}^{\dagger}c_j$, 
is found by choosing $H=-J\sum_j\left(c^{\dagger}_{j+1}c_{j}+H.c.\right)$ and the non-local jump operators $L_j=\sqrt{\kappa}\left(c_j\pm ic_{j+1}\right)$ \cite{Gong2017Zeno}.
Nonlocal one-body loss terms $L_j$
can be obtained by adiabatically eliminating a fast-decaying internal excited state in a 
(anti-)magic wavelength in alkaline earth atoms \cite{Reiter2012,Gong2018}.

This work opens up new possibilities for future investigations in Anderson localized systems. 
For example, we show that the absence of the boomerang effect can be used to find transitions to phases with extended states or to phases with complex eigenenergies.
The models studied here may be implemented using cold atoms in optical lattices and the quantum boomerang effect could be experimentally verified in such effective non-Hermitian systems. 
Future possible investigations include {\it (i)} an understanding of why the boomerang effect may present oscillations in some models with pseudodisorder; {\it (ii)} a better comprehension of the second return presented in models with complex spectrum; and {\it (iii)} a derivation of a generalized dynamical relation valid for non-Hermitian systems.
Finally, a relevant open question concerns the fact of whether or not certain interactions preserve the quantum boomerang effect. Specifically, more sofisticated numerical methods can be used to investigate the presence of the quantum boomerang effect in the context of many-body localized phases.

\section*{Acknowledgements}
We acknowledge P. Vignolo and L. Tessieri for useful feedback on the manuscript and A. Moustaj, D. P. Pires and C. Gao for helpful discussions. 
We thank the High-Performance Computing Center (NPAD) at UFRN for providing computational resources. T. M. acknowledges the hospitality of ITAMP-Harvard where part of this work was done.
T.M.~ acknowledges CNPq for support through 
Bolsa de produtividade em Pesquisa. T.~M. and F.~N. acknowledge support from CAPES.
This work was supported by the Serrapilheira Institute 
(grant number Serra-1812-27802). 


\begin{appendix}

\section{Discussion on the Hatano-Nelson model}\label{app.A}

\begin{figure}[b]
\centering 
\includegraphics[width=\columnwidth]{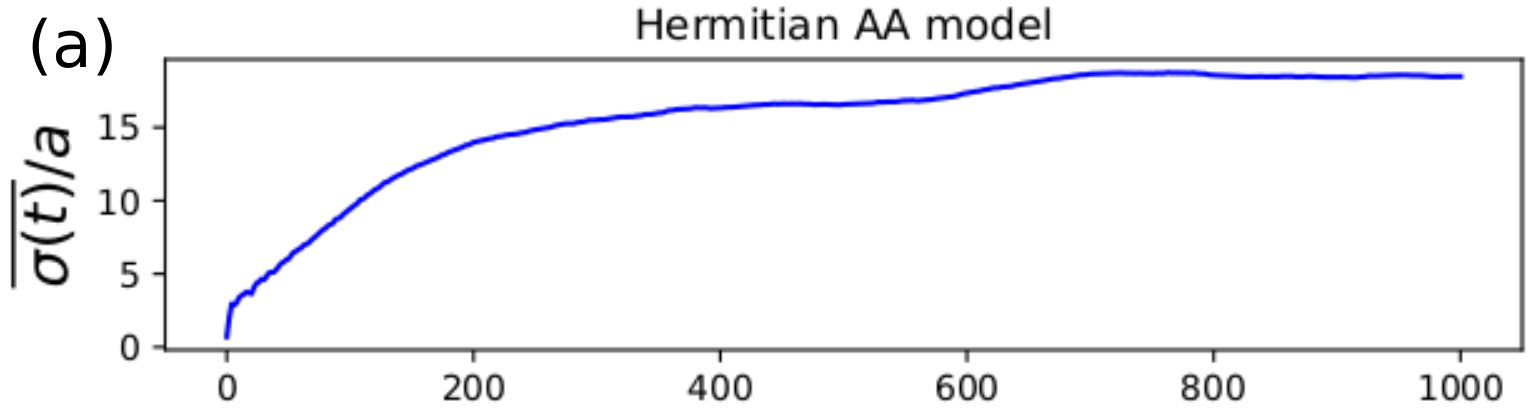}\vspace{0.2cm}
\includegraphics[width=\columnwidth]{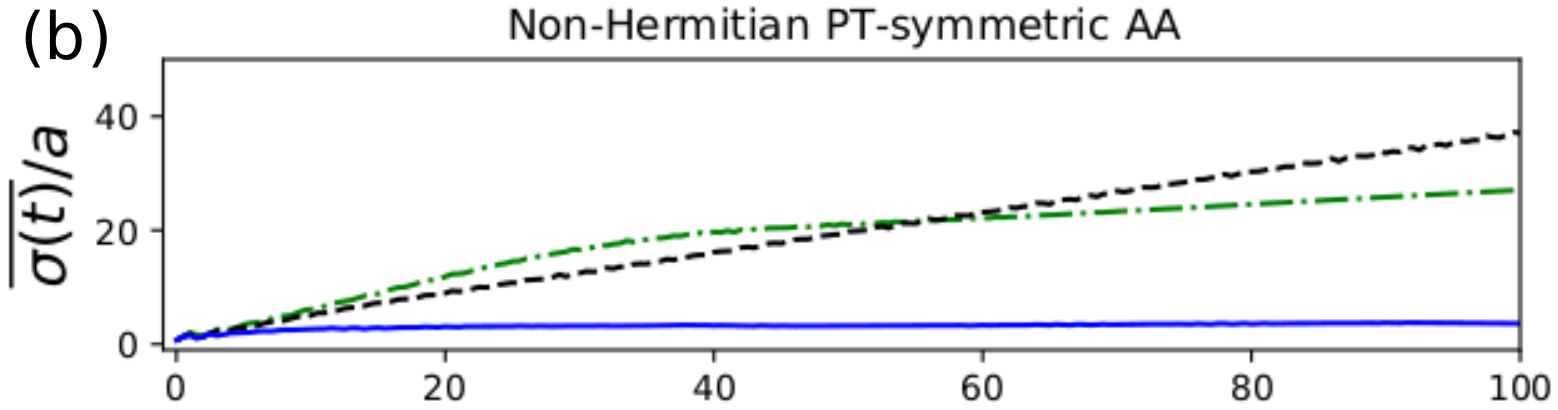}\vspace{0.2cm}
\includegraphics[width=\columnwidth]{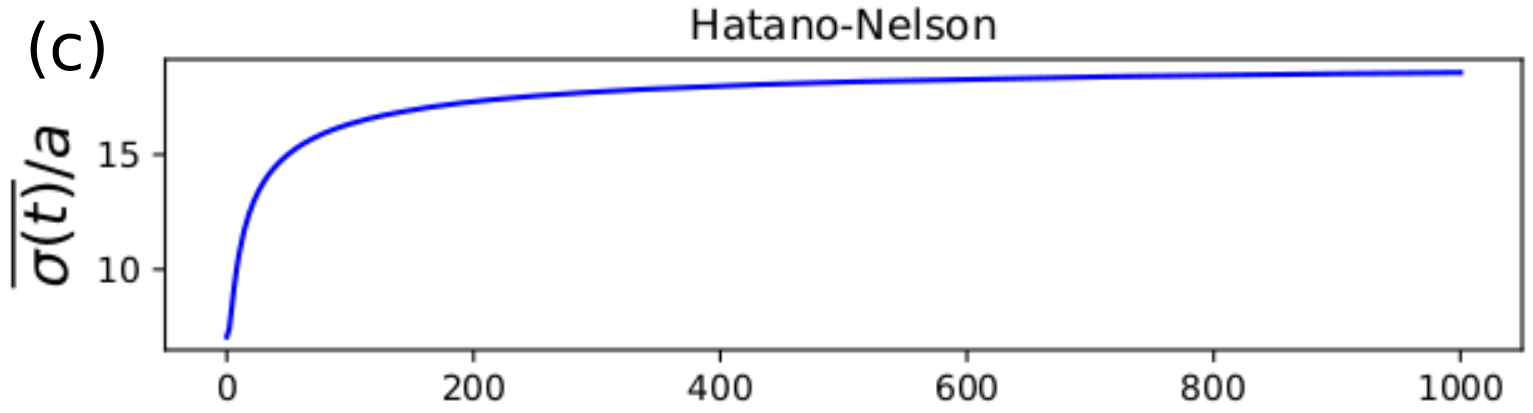}\vspace{0.2cm}
\includegraphics[width=\columnwidth]{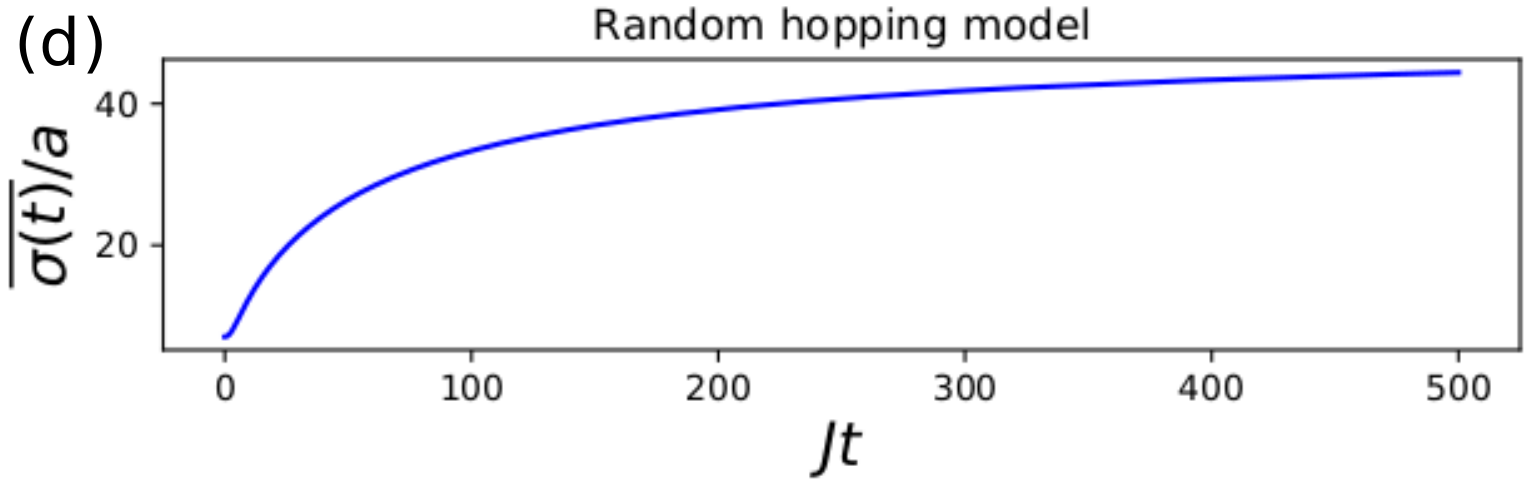}
\caption{\textbf{Variance $\overline{\sigma(t)}$.} 
(a) Hermitian Aubry-Andr\'e model using the same parameters of Fig.~\ref{AA_model}(a). 
(b) Non-Hermitian Aubry-Andr\'e model using the same parameters of Figs.~\ref{AA_model}(b)-(d) with $W/J=1.6$ in green dot-dashed (extended \textit{PT}-broken phase), $W/J=1.8$ in black dashed (extended \textit{PT}-symmetric phase) and $W/J=2.2$ in blue solid line (localized \textit{PT}-symmetric phase), respectively.
(c) Hatano-Nelson model using the same parameters of Fig.~\ref{boomerang_HNa}(a). 
(d) \textit{T}-symmetric non-Hermitian random hopping model using the same parameters of Fig.~\ref{randomhop}.
} \label{variance}
\end{figure}

\begin{figure}[b]
\centering 
\includegraphics[width=\columnwidth]{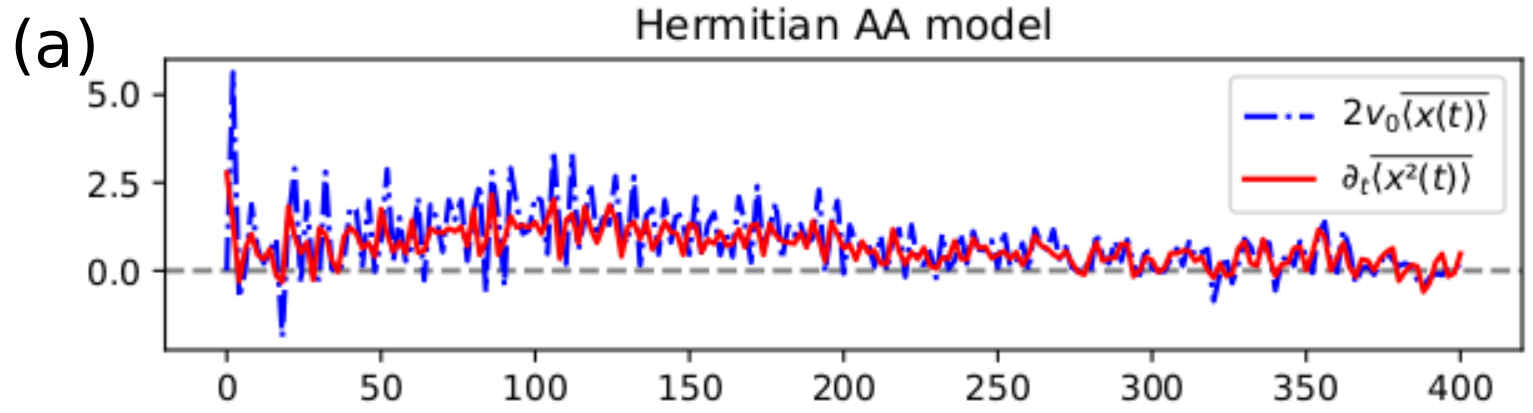}\vspace{0.2cm}
\includegraphics[width=\columnwidth]{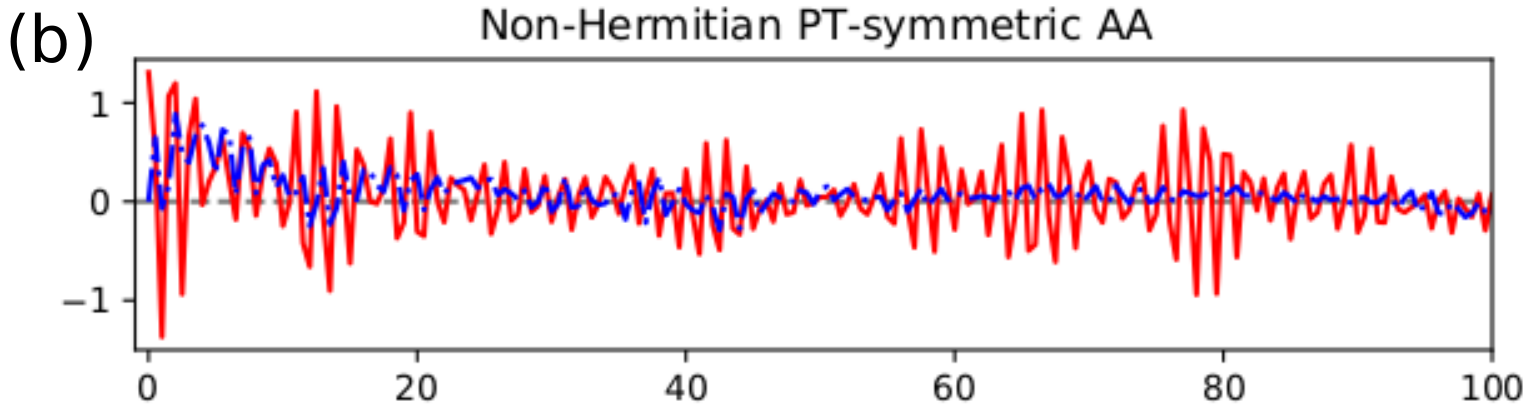}\vspace{0.2cm}
\includegraphics[width=\columnwidth]{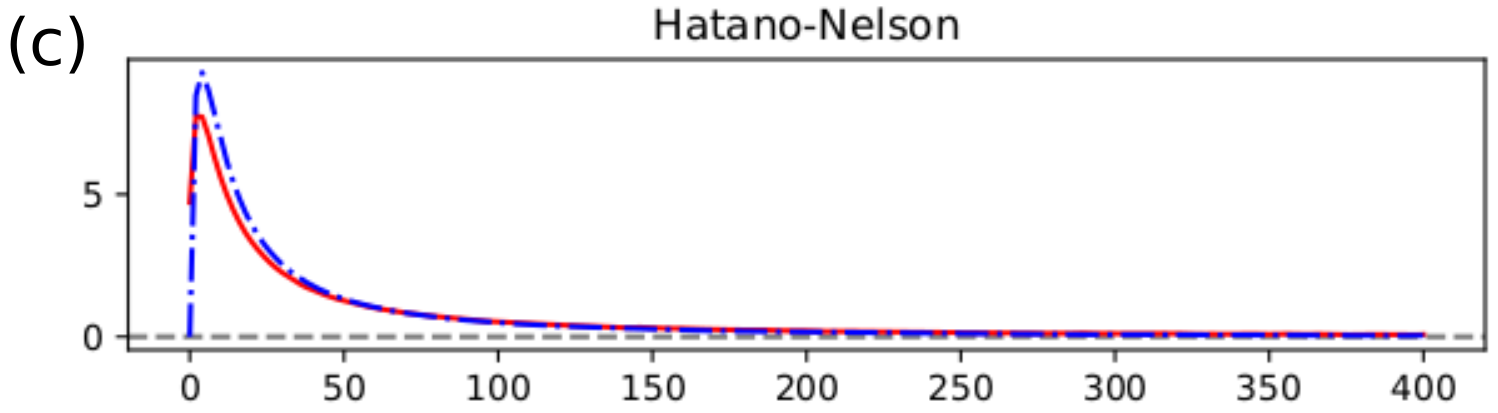}\vspace{0.2cm}
\includegraphics[width=\columnwidth]{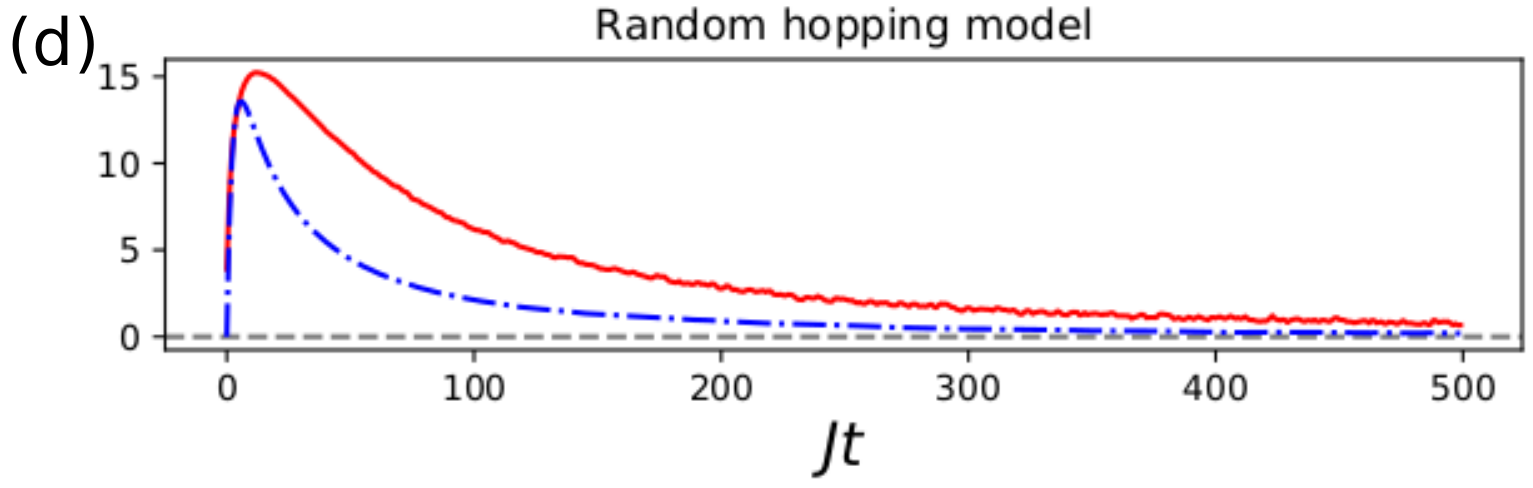}
\caption{\textbf{Breakdown of the dynamical relation.} 
(a)~Hermitian Aubry-Andr\'e model using the same parameters of Fig.~\ref{AA_model}(a). 
(b) Non-Hermitian Aubry-Andr\'e model using the same parameters of Fig.~\ref{AA_model}(d) (localized \textit{PT}-symmetric phase).
(c) Hatano-Nelson model using the same parameters of Fig.~\ref{boomerang_HNa}(a). 
(d) Non-Hermitian random hopping model using the same parameters of Fig.~\ref{randomhop}.
} \label{magic}
\end{figure}

We numerically check that in the standard HN model if the disorder strength $W$ is such that all the states are localized and if the initial momentum is zero, $k_0=0$, then $\overline{\langle x (t) \rangle_>}$ $\left(\textrm{or } \overline{\langle x (t) \rangle_<}\right)$ is a monotonically non-decreasing (or non-increasing) function of $t$ that tends to a finite value when $t\to+\infty$.
This happens because all the disorder realizations in $\overline{\langle x (t) \rangle_>}$ are such that the hopping to the right is larger than the hopping to the left and hence the center of mass have a tendency to move to the right. The reverse is true for $\overline{\langle x (t) \rangle_<}$ and the center of mass tends to move to the left.
More generally, for any initial momentum $k_0$ one has finite values for the center of mass at asymptotically long times, i.e.~$\lim_{t\to+\infty} \overline{\langle x (t) \rangle_>}>0$ and $\lim_{t\to+\infty} \overline{\langle x (t) \rangle_<}<0$, though for finite $k_0$ the functions 
$\overline{\langle x (t) \rangle_\lessgtr}$
are not necessarily monotonic functions of the time $t$ and they may present a U-turn in some cases [see Fig.~\ref{boomerang_HNa}(a)].  In the localized phase of the HN model we find that the asymptotic values $\lim_{t\to+\infty} \overline{\langle x (t) \rangle_\lessgtr}$ depend on $|k_0|$, though they do not depend on $\textrm{sign}(k_0)$. The choice of $k_0$ may also affect the value of $\overline{\langle x (t_U) \rangle_\lessgtr}$ and the height of the U-turn $\overline{\langle x (t_U) \rangle}$.

\section{Variance saturation and breakdown of dynamical relation}\label{app.B}

Generally speaking, the QBE takes place in Anderson-localized systems and this localization leads to the absence of diffusion. Therefore, usually the variance of the wave packet is bounded. However, at the critical point $W_c$ of the three-dimensional Anderson model the size of the wave packet increases like $t^{1/3}$ and the QBE still appears~\cite{Prat2019Quantum}. 
In Fig.~\ref{variance} we compute the variance using $\overline{\sigma(t)}^2=\overline{<x^2(t)>}-\overline{<x(t)>}^2$ for several models considered in this paper that present the QBE, namely the Hermitian AA model [panel (a)], the localized PT-preserved phase of the non-Hermitian AA model [panel (b)], the HN model [panel (c)] and the \textit{T}-symmetric random hopping model with real spectrum [panel (d)]. In all these cases the variance $\overline{\sigma(t)}$ saturates as $t\to+\infty$. This signalizes Anderson localization in the investigated models.
We also show in panel (b) the variance in the non-Hermitian AA model in the extended \textit{PT}-symmetric phase and in the extended \textit{PT}-broken phase. The variance has a different behavior in each of the three phases of the non-Hermitian \textit{PT} symmetric AA model.

In Fig.~\ref{magic} we show the breakdown of the dynamical relation Eq.~(\ref{dynamical}), derived for Hermitian models, where $v_0=2Ja\, \textrm{sin}(k_0a)$ is computed from the disorderless model~\cite{Prat2019Quantum,Tessieri2021Quantum}. In Fig.~\ref{magic}(a) we consider the standard AA model. As it is an Hermitian model, the dynamical relation holds reasonably well, though there are lots of fluctuations. In Fig.~\ref{magic}(b) the non-Hermimtian AA model is considered in the localized \textit{PT}-symmetric phase. The agreement between $\partial_t \overline{\langle x^2(t) \rangle}$ and $2v_0\overline{\langle x (t)\rangle}$ is visibly worse in this case. In Fig.~\ref{magic}(c) we consider the HN model with a small non-Hermiticity $|J_a|/J=0.01$. In this case the dynamical relation can be a reasonable approximation. However, increasing $|J_a|/J$ leads to the breakdown of the dynamical relation. Figure~\ref{magic}(d) shows the breakdown of that relation for the \textit{T}-symmetric non-Hermitian random hopping model in the phase with real spectrum. We note that no constant factor $\beta$ could be used in order to generalize the dynamical relation in the form $\partial_t \overline{\langle x^2(t) \rangle}=2\beta v_0\overline{\langle x(t) \rangle}$ for the non-Hermitian models. 

\end{appendix}

\bibliography{Reference}

\begin{thebibliography}{66}
\bibitem [{foo()}]{footeq2}%
  \BibitemOpen
  \href@noop {} {\bibinfo {title} {For each disorder realization one has that
  $\langle x(t)\rangle=\sum_{n\neq m} c_n c_m^*\,\textrm{exp}[-i(\epsilon_n -
  \epsilon_m)t]\bra{\phi_m}x \ket{\phi_n}+\sum_n\left|c_n\right|^2\bra{\phi_n
  }x\ket{\phi_n}$ presents fluctuations with time due to the off-diagonal terms
  and they oscillate about a time-independent component given by the diagonal
  terms. {It} is assumed that performing the average over disorder realizations
  at long times causes these time-dependent terms to cancel due to the random
  phases. {We} numerically confirm the validity of the diagonal ensemble due to
  the disorder average.}}\BibitemShut {Stop}%
\end{thebibliography}


\end{document}